\documentclass[a4paper,fleqn]{cas-dc}

\usepackage[authoryear]{natbib}
\defcitealias{cho2014learning}{Cho et al., 2014}

\usepackage{amsmath,amssymb,amsthm}
\usepackage{graphicx}
\graphicspath{{figures/}}
\DeclareGraphicsExtensions{.pdf,.png,.jpg}
\usepackage{booktabs}
\usepackage{array}
\usepackage{tabularx}
\usepackage{algorithm}
\usepackage{algpseudocode}
\usepackage{multirow}
\usepackage{placeins}
\newtheorem{theorem}{Theorem}[section]
\newtheorem{definition}{Definition}[section]
\newtheorem{proposition}{Proposition}[section]

\begin{document}
\let\WriteBookmarks\relax
\makeatletter
\setlength{\@fptop}{0pt}
\setlength{\@fpsep}{8pt plus 2pt minus 2pt}
\setlength{\@fpbot}{0pt plus 1fil}
\setlength{\@dblfptop}{0pt}
\setlength{\@dblfpsep}{8pt plus 2pt minus 2pt}
\setlength{\@dblfpbot}{0pt plus 1fil}
\makeatother
\def\textpagefraction{.001}
\setcounter{topnumber}{5}
\setcounter{dbltopnumber}{4}
\setcounter{totalnumber}{10}
\renewcommand{\topfraction}{0.95}
\renewcommand{\dbltopfraction}{0.95}
\renewcommand{\textfraction}{0.05}
\renewcommand{\floatpagefraction}{0.75}
\renewcommand{\dblfloatpagefraction}{0.75}

\shorttitle{Safety-Contract Graph MARL}
\shortauthors{Jose Luis Silva}

\title[mode=title]{Safety-Contract Graph Multi-Agent Reinforcement Learning for
  Autonomous Network Security Response}


\author[1,2]{Jose Luis Lima de Jesus Silva}
\ead{jseluis.silva@gmail.com}
\ead[url]{https://oxaala.com.br}
\credit{Conceptualization, Methodology, Software, Formal analysis,
  Investigation, Data curation, Writing the original draft, review \& editing, and Visualization}

\affiliation[1]{
  organization={Oxaala Tecnologias},
  addressline={Rua Dinah Silveira de Queirós, 06, Quinta do Candeal, Horto Florestal},
  city={Salvador},
  postcode={40296-160},
  state={Bahia},
  country={Brazil}
}
\affiliation[2]{
  organization={Universidade Federal da Bahia},
  addressline={Instituto de Geociências, Rua Barão de Jeremoabo, s/n, Ondina},
  city={Salvador, Bahia},
  postcode={40170-115},
  country={Brazil}
}

\cortext[1]{Corresponding author}
\begin{abstract}
Autonomous network-security response systems promise to reduce Security
Operations Centre (SOC) reaction latency, but reward-only multi-agent
reinforcement learning (MARL) can improve security reward while remaining
non-deployable. We present a safety-contract graph MARL framework and
instantiate it as ACD$^3$-GAT (Adaptive Constrained Counterfactual Decisioning
with a Graph Attention Network encoder), an architecture that separates
simulator observations from reusable operational budgets, constrained
optimization, graph state encoding, and counterfactual action screening. We
evaluate the method in CAGE Challenge~4, where agents operate under budgets
for Mean Time to Recover (MTTR), false-positive response, and firewall
change-management disruption. Across the benchmark, every unconstrained method
violates the SOC downtime budget in 100\% of evaluated episodes, with mean
downtime proxy costs of 311--430 against a budget of 50. This complements
prior CAGE Challenge~4 findings by showing that reward-only learning lacks
operational discipline. Constrained MAPPO-GAT (C-MAPPO-GAT) isolates
Lagrangian operational-cost control and budget-aware screening, while
ACD$^3$-GAT adds budget context, CVaR tail-risk estimation, opponent-belief
state, and Graph Counterfactual Risk Propagation (G-CRP). The replicated
comparison includes three 200-episode seeds for IPPO, MAPPO-GAT, C-MAPPO-GAT,
and ACD$^3$-GAT. C-MAPPO-GAT reduces downtime violation from 100\% to 0.3\%
and mean downtime cost from 355.4 to 15.5 relative to MAPPO-GAT.
ACD$^3$-GAT reduces mean downtime cost to 48.2 with a 13.8\% violation rate,
placing it on the safety-contract frontier rather than at the most
conservative compliance point. Topology-seed and coupled adaptive Red-process
stress tests preserve this contrast and show lower worst adaptive degradation
for safety-constrained policies than reward-only MAPPO-GAT.
\end{abstract}




\maketitle


\section{Introduction}
\label{sec:intro}

The volume and velocity of network-security incidents facing modern enterprises have outpaced human response capacity~\citep{vyas2023acd}.
Security Operations Centers (SOCs) must triage thousands of alerts
daily, manually correlating events, isolating hosts, and reimaging compromised systems, and adjusting firewall policies while maintaining the availability of mission-critical services. Intelligent network security agents that can recommend, screen, or execute these responses are therefore an active and growing research priority for Civilian SOC automation and enterprise network management. This setting is intrinsically structured and sequential: in CAGE Challenge~4, five CAGE ``Blue'' agents act over a 500-step episode from partial binary observations of a changing enterprise network, choosing action types and topology-dependent targets while the simulator's CAGE ``Red'' process continues to discover, escalate, and disrupt services. A response action can have delayed, competing effects, whether it is a Restore, which may remove a compromised session while taking a host offline, or \textit{BlockTrafficZone}, which may protect one mission path while disrupting another. Furthermore, a missed response may only become visible many steps later. Therefore, an autonomous network security response requires long-horizon multi-agent coordination that accounts for partial observability, invalid-action constraints, and non-stationarity due to opposing processes. The learning objective must therefore capture both containment of malicious activity and the operational acceptability of the resulting intervention pattern.
Existing autonomous cyber-response research using reinforcement learning (RL) typically optimizes a scalar security reward without imposing explicit operational constraints. In practice, every response action carries an operational cost. For example, a host \textsc{Restore} action takes the system offline for the duration of reimaging.

Repeated \textsc{BlockTrafficZone} and \textsc{AllowTrafficZone} commands create firewall-policy violations under change-management governance, while \textsc{Restore} actions issued without clear malicious evidence constitute false-positive responses that burden SOC analysts. If an RL agent learns to issue excessive Restore actions to clear compromised sessions, which is a sensible strategy for maximizing security reward in the simulator, it may simultaneously exhaust the organization's Mean Time to Recover (MTTR) budget, flood the change management system, and generate analyst fatigue.

In our evaluation, every unconstrained learning method reaches a
downtime-budget violation rate of 100\% ($P(\mathrm{viol}^{\mathrm{DT}})=1.00$), with mean per-episode downtime proxy costs of 311--430 Restore-action events against an episode budget of 50. This pattern complements the central finding of the CAGE Challenge~4 evaluation~\citep{kiely2025cage4}, where carefully engineered heuristics outperformed the submitted multi-agent reinforcement learning (MARL) agents because they encoded valid-action handling, mission-phase traffic discipline, and selective incident-response logic that reward-only MARL did not reliably discover.

Our contribution is to make that discipline explicit as a measurable safety
contract rather than leaving it implicit in manually engineered policies. As a
direct consequence, the learned agent becomes an auditable decision-support
component whose actions can be checked against local MTTR, firewall-change,
and false-positive response budgets.

We also introduce \textbf{ACD$^3$-GAT} (Adaptive Constrained Counterfactual
Decisioning with a Graph Attention Network), a safety-contract graph MARL framework for autonomous network security response. The framework builds on established components used in the CAGE cyber-range evaluation~\citep{cage4,kiely2025cage4}, Graph Attention Networks~\citep{velickovic2018graph,sandoval2025attentive},
Proximal Policy Optimization (PPO) and MAPPO-style multi-agent policy
optimisation~\citep{schulman2017proximal,yu2022surprising}, graph-based
generalisation~\citep{king2025graphacd}, and constrained reinforcement
learning~\citep{altman1999constrained,achiam2017constrained}. Our contribution is to combine them around a specific deployability problem, where the SOC operational budgets must be represented explicitly during training, action screening, and evaluation. In this formulation, the downtime/MTTR, false-positive response, and firewall change disruptions are not treated as incidental side effects of the reward function, but as operational quantities that the policy must learn to respect and that the evaluator can audit.

We introduce C-MAPPO-GAT as a controlled constrained instantiation that combines a MAPPO centralized critic, a GAT observation encoder, Lagrangian operational-cost advantages, and a budget-exhaustion fallback under the same SOC contract. This is a baseline that isolates the effect of adding explicit operational-cost control before the broader ACD$^3$-GAT architecture adds budget context, tail-risk estimation, opponent-state information, and counterfactual action screening.

The framework separates reusable safety-contract components from
simulator-specific interfaces. Operational budgets, Lagrangian cost learning, graph-structured encoding, counterfactual screening, and tail-risk accounting form the reusable safety-contract machinery. By contrast, the observation parser and valid-action mapping are treated as environment interfaces. The empirical claims in this paper are therefore kept deliberately grounded in the CAGE-4 evaluation and the reported robustness stress tests.

Autonomous network-security response is treated here as a constrained Decentralized Partially Observable Markov Decision Process (Dec-POMDP). The formulation optimizes response policies not only for CAGE-4 reward, but also for operational acceptability under SOC budget constraints. We therefore attach three budget counters to the response process: (i) \textsc{Restore} action is counted against the downtime and MTTR budget ($B_{\mathrm{down}}{=}50$), (ii) a response taken without visible malicious evidence is counted against the false-positive budget, which represents analyst burden ($B_{\mathrm{fp}}{=}10$), and (iii) traffic-control actions are counted against the firewall-disruption budget, which represents change-management pressure ($B_{\mathrm{fw}}{=}20$). The same counters are used during training, screening, and evaluation, and an episode is acceptable only if the agent contains the simulated intrusion without exceeding any SOC budget.

The framework turns this contract into an expert-system layer for graph MARL. Operational budgets, Lagrangian cost learning, action-screening rules, and safety-labelled trajectory logging are treated as reusable decision-support components, while the CAGE-4 observation parser and valid-action mapping remain simulator-specific interfaces. This separation keeps the empirical claims grounded in CAGE-4 while allowing the contract layer to be reused with other observation formats and action schemas.

Our resulting architecture, the ACD$^3$-GAT, combines the host-subnet graph encoders, a factorized target-action policy, operational cost critics, CVaR-based tail-risk estimation, override signals, Lagrangian cost learning, and Graph Counterfactual Risk Propagation (G-CRP). 

We also introduce the C-MAPPO-GAT as a controlled, constrained graph-MARL baseline, which isolates the effect of explicit operational-cost control before the broader ACD$^3$ architecture adds budget context, opponent-state information, tail-risk accounting, and counterfactual action screening.

In this work, we evaluate reward-only, graph-attentive, Lagrangian-constrained, and ACD$^3$ policies under the same safety-contract metrics, with explicit seed and episode reporting. We also introduce Temporal Contract Graph Shielding (TCGS) is evaluated as a diagnostic extension that estimates whether a candidate action is likely to push the episode beyond a SOC budget given the recent history of alerts, actions, rewards, and accumulated costs. TCGS learns this future budget-violation risk from safety-labelled trajectory histories and uses the prediction to screen frozen ACD\textsuperscript{3}-GAT action proposals before execution. It is therefore reported as evidence that temporal contract-risk prediction can support action screening, rather than as a claim that we have trained a new end-to-end TCGS policy.

Experiments on CAGE Challenge~4 show that the replicated core benchmark has three 200-episode seeds for Independent Proximal Policy Optimisation (IPPO), Multi-Agent PPO with a GAT encoder (MAPPO-GAT), constrained MAPPO-GAT, and ACD$^3$-GAT. 

The results show that the Constrained MAPPO-GAT reduces downtime violation from 100\% to 0.3\% and mean downtime cost from 355.4 to 15.5 relative to MAPPO-GAT. ACD\textsuperscript{3}-GAT reduces mean downtime cost to 48.2 with a 13.8\% violation rate, placing the integrated method on a broader safety-contract frontier rather than at the most conservative compliance point.

This result separates two claims: ACD\textsuperscript{3}-GAT defines the general safety-contract architecture, while the experiments identify which configurations already deliver reliable operational compliance.

We have also conducted robustness stress tests, which strengthen this interpretation, since the constrained MAPPO-GAT preserves the downtime contract under topology-seed shifts, and the coupled adaptive Red-process stress test shows lower worst-case degradation for constrained and ACD$^3$ policies than for reward-only MAPPO-GAT.


\section{Related Work}
\label{sec:related}

\subsection{Autonomous Network Security Environments}

The growth of autonomous cyber-response research has driven a proliferation of simulation environments~\citep{vyas2023acd}. CybORG~\citep{standen2021cyborg} underpins the CAGE challenge series and provides the enterprise network model used in this work. The CAGE Challenge 4 is the first multi-agent CAGE variant, requiring five
cooperating CAGE ``Blue'' response agents to operate against a persistent
CAGE ``Red'' process across 500-step episodes~\citep{cage4,kiely2025cage4}.
The principal finding of the CAGE~4 evaluation, that carefully engineered
heuristics outperformed the submitted multi-agent reinforcement learning (MARL) agents, motivating our focus on operationally constrained training.
Those heuristics encoded valid-action handling, event filtering, and mission-phase traffic discipline, and selective Restore/Remove behavior that reward-only MARL agents did not learn reliably. Beyond CAGE studies have also shown that Proximal Policy Optimisation (PPO)-family response policies can degrade under unseen networks and opposing strategies~\citep{wolk2022beyond}. Together, these results motivate a deployability question that is distinct from reward generalisation, and focus on whether a learned responder remains within the operational budgets that a SOC would impose on its actions.

\subsection{MARL for Network Security Response}

Multi-agent reinforcement learning has been applied to network intrusion
response~\citep{vyas2023acd}, with most prior work using independent
actor-critic variants~\citep{de2020independent} or centralised
critics~\citep{yu2022surprising}.
The published CAGE~4 analysis reports that the best default CAGE Challenge 4 (CC4) top-team entry was heuristic ($-113 \pm 35$ mean return) while the top-team MARL entry scored $-193 \pm 84$ over 100 episodes of 500 steps~\citep{kiely2025cage4b}. The $-101 \pm 36$ value is the constant-network-size reference setting, which is useful as a reference point but not the default CC4 score~\citep{kiely2025cage4b}.
Our work addresses the same gap from a deployability perspective by imposing explicit operational constraints. Recent Large Language Model (LLM)-based CAGE~4 agents achieve approximately $-2{,}888$
with role prompting (GPT-o1-mini)~\citep{castro2025llm}.
Hierarchical MARL for CAGE~4 decomposes the response into sub-policies for investigation and recovery, improving convergence and reporting interpretable metrics such as clean-machine ratio, precision, and false
positives~\citep{singh2024hierarchical}. A recent LLM-based CAGE~4 work instead emphasizes explainability, natural-language observation formatting, and communication among LLM/RL teams~\citep{castro2025llm}. These lines complement our contribution because they improve task decomposition, reasoning, or communication, whereas ACD\textsuperscript{3}-GAT turns operational harms into explicit episode-level SOC budgets and optimizes or screens policies against those budgets.

\subsection{Constrained and Safe Reinforcement Learning}

The Constrained MDPs~\citep{altman1999constrained} formalize budget constraints via Lagrangian relaxation, with theoretical convergence results under standard regularity conditions. Furthermore, the Constrained policy optimization extends this idea to deep policy gradient
settings with trust-region style constraint handling~\citep{achiam2017constrained}.
Our neural MARL setting is non-convex, so we use this machinery as a
practical budget-enforcement mechanism rather than as a proof of formal
constraint satisfaction. Additionally, Safe reinforcement learning (RL) benchmarks evaluate single-agent constraint satisfaction~\citep{ray2019benchmarking}.
Therefore, we extend this to multi-agent network-security response with three simultaneous SOC constraints. Our Lagrangian update (dual step size $\eta_\lambda{=}0.01$) follows the application of projected subgradient methods on the dual variables.

\subsection{Graph Neural Networks in Security}

Graph-structured representations have been applied to network intrusion
detection and malware classification, often through message-passing neural
networks~\citep{gilmer2017neural} and inductive neighborhood aggregation
such as GraphSAGE~\citep{hamilton2017inductive}.
Temporal graph networks provide a general memory-based framework for dynamic graphs~\citep{rossi2020temporal}, and cyber detection systems have used spatio-temporal graph neural networks (GNNs) for smart-grid intrusion localization~\citep{haghshenas2022tgnn} and network-intrusion
detection~\citep{vanlangendonck2024pptgnn}.

Those works primarily solve detection, localization, or dynamic-graph
prediction problems. Within autonomous network response, attentive graph agents have already shown that Graph Attention Network (GAT) policies can exploit network topology and adapt across changed network structures~\citep{sandoval2025attentive}, while graph-based RL agents represent ACD observations and actions as attributed graphs to improve zero-shot topology generalisation~\citep{king2025graphacd}.

Accordingly, graph attention is the perception layer in our system, while the new problem formulation is a safety-contract graph MARL, where the key evaluation question is whether response actions remain within explicit SOC budgets. This positioning also clarifies the role of C-MAPPO-GAT in our benchmark. The MAPPO~\citep{yu2022surprising}, GAT encoders~\citep{velickovic2018graph}, and constrained MDP/Lagrangian safety methods \citep{altman1999constrained,achiam2017constrained}
are established components, but C-MAPPO-GAT is introduced here as their
controlled SOC-budget instantiation for CAGE-4 response. It is therefore not treated as a named prior algorithm; it is the constrained baseline that isolates the effect of explicit operational costs before the broader ACD\textsuperscript{3}-GAT architecture adds budget context, tail-risk accounting, opponent state, and counterfactual screening.

\begin{table*}[!t]
\centering
\caption{Positioning relative to closely related autonomous cyber-defence
research.  The comparison emphasises the evaluation axis rather than ranking
prior systems by return or topology generalisation.  Citations in the first
column identify representative work for each research line.}
\label{tab:related_positioning}
\begingroup
\footnotesize
\setlength{\tabcolsep}{3.2pt}
\renewcommand{\arraystretch}{1.12}
\begin{tabularx}{\textwidth}{@{}>{\raggedright\arraybackslash}p{0.22\textwidth}
>{\raggedright\arraybackslash}p{0.31\textwidth}
>{\raggedright\arraybackslash}X@{}}
\toprule
Research line & Primary objective & Deployability axis addressed here \\
\midrule
CAGE/CybORG benchmarks and Beyond CAGE~\citep{standen2021cyborg,cage4,kiely2025cage4,wolk2022beyond}
& Cyber-range evaluation, reward generalisation, and comparison with engineered
heuristics
& We add explicit episode-level SOC budgets and violation rates, showing that
reward-improving MARL can remain operationally non-deployable. \\
\addlinespace
Hierarchical and LLM-assisted CAGE-4 agents~\citep{singh2024hierarchical,castro2025llm}
& Task decomposition, convergence, clean-machine ratio, precision/false
positives, reasoning, and communication
& We treat downtime, false-positive response, and firewall-change disruption as
governance constraints optimized or screened during response. \\
\addlinespace
Topology-adaptive GAT and graph-RL defenders~\citep{velickovic2018graph,sandoval2025attentive,king2025graphacd}
& Graph representation, topology adaptation, and zero-shot generalisation of
cyber-defence policies
& Graph attention is used as the perception layer inside a constrained
safety-contract policy rather than as the sole novelty or the headline claim. \\
\addlinespace
Temporal graph and cyber-detection GNNs~\citep{rossi2020temporal,haghshenas2022tgnn,vanlangendonck2024pptgnn}
& Dynamic graph memory, intrusion detection, and attack localisation
& The response problem adds multi-agent action selection and operational costs;
TCGS is a diagnostic step toward temporal contract-risk screening. \\
\addlinespace
Asynchronous cyber-range MARL~\citep{jankowski2026netforge}
& Simulator realism, continuous-time telemetry, and Sim2Real evaluation
& The proposed safety-contract layer can be ported to richer environments once
their action traces support SOC cost accounting. \\
\bottomrule
\end{tabularx}
\endgroup
\end{table*}

\subsection{Risk-Sensitive Reinforcement Learning}

Conditional Value-at-Risk (CVaR) optimisation targets the tail of
the return distribution~\citep{rockafellar2000optimization}.
This is particularly important in cybersecurity, where a single catastrophic episode (full critical-zone compromise) may outweigh many successful ones. We integrate CVaR episode reweighting into the MARL training loop via batch-level importance weights.

\subsection{Adaptive Opposing-Policy Evaluation as Robustness Context}

Adaptive opposing-policy evaluation is a standard way to probe whether an
autonomous response policy is brittle to changes in the simulated source of malicious activity, but it is rarely formalised in CAGE-style benchmarks. We retain this as a robustness extension in the evaluation protocol, closest in spirit to self-play in multi-agent game-playing AI~\citep{bansal2018emergent} and population-based training
\citep{jaderberg2019human}, adapted to the asymmetric CAGE Blue--Red setting of enterprise network-security response. Recent asynchronous cyber-range work argues that simulator-to-SOC transfer also requires continuous-time events, noisy telemetry, and richer hypervisor-backed evaluation~\citep{jankowski2026netforge}.
Our contribution is orthogonal to that simulator-realism direction, as the safety-contract layer introduced here can be instantiated in CAGE-4 today and can also serve as the governance layer for future temporal or asynchronous cyber-range environments.


\section{Evaluation Scope and Operational Safety Problem}
\label{sec:threat}

\subsection{Network and Asset Model}

We adopt the CAGE Challenge~4 enterprise network
model~\citep{cage4,kiely2025cage4}.
The network comprises nine subnets spanning internet-facing, contractor,
office, administrative, restricted, and operational zones.
Mission-critical assets reside in two ``operational zones'' whose compromise
or unavailability directly degrades the simulated mission score.

All response actions are abstract simulation primitives, where no exploit
code, credentials, malware, vulnerability details, or real network
infrastructure are involved. The Red process's actions (discover, exploit, escalate, degrade, impact, withdraw) are likewise simulator-internal operations. This paper is therefore a civilian enterprise SOC decision-support study conducted inside a closed simulator, and it provides no real-world intrusion capability.

\subsection{Opposing Process Model}

The CAGE Red process begins with a session on the contractor subnet (simulating a
supply-chain foothold) and attempts lateral movement toward restricted and
operational zones.
We evaluate against three Red-process strategy classes:

\begin{enumerate}
\item \textbf{Finite-state machine (FSM) Red process}: the default CAGE-4 scripted process
  that selects targets probabilistically using a fixed host-state transition
  matrix~\citep{kiely2025cage4}.
\item \textbf{Discovery Red}: a variant that emphasises reconnaissance before
  impact, creating longer dwell time.
\item \textbf{Learned Red (adaptive)}: a PPO-trained Red process adapted online
  against a frozen Blue policy, used exclusively for exploitability evaluation.
\end{enumerate}

\subsection{Operational Safety: SOC Budget Constraints}

A response policy is not considered operationally successful merely by achieving
high security reward.
Real SOC governance imposes three categories of operational constraint that
map directly to our cost signals (Section~\ref{sec:costs}):
The budgets are deliberately explicit stress-test thresholds for the CAGE-4
simulator rather than universal SOC constants; in deployment they would be set
by local service-level objectives, change-control policy, and analyst
capacity.

\paragraph{(1) Mean Time to Recover (MTTR) --- $c_t^{\mathrm{down}}$.}
Host reimaging (Restore) renders a machine unavailable for several minutes
in practice.
Repeated reimaging of hosts violates MTTR service-level agreements (SLAs)
and degrades green-user access to services.
We proxy this with a count of Restore actions per episode, budgeted at
$B_{\mathrm{down}}{=}50$.

\paragraph{(2) False-Positive Response Rate --- $c_t^{\mathrm{fp}}$.}
Restore or Remove actions issued in the absence of clear malicious evidence
constitute false-positive responses.
They consume analyst time for post-action review, create unnecessary
disruption, and erode operator trust in the autonomous system.
Budget: $B_{\mathrm{fp}}{=}10$.

\paragraph{(3) Firewall Change-Management Policy --- $c_t^{\mathrm{fw}}$.}
BlockTrafficZone and AllowTrafficZone actions modify network topology.
In enterprise environments, such changes require change-management approval
and create audit trails.
Excessive firewall churn violates change-management governance.
Budget: $B_{\mathrm{fw}}{=}20$.

\paragraph{Safety contract.}
A CAGE Blue policy \emph{satisfies the safety contract} in an episode iff
$\sum_t c_t^k \leq B_k$ for all three constraints.
The violation rate $P(\mathrm{viol}^k)$ is the primary safety metric:
a deployed autonomous response system must meet a pre-specified violation tolerance with
confidence intervals, audit logs, and an escalation path for budget exhaustion.
We therefore treat lower violation rate and tighter confidence bounds as
deployability evidence rather than claiming formal certification or a
mathematical safety guarantee for non-convex neural policies.

\paragraph{Ethical statement.}
All experiments are conducted entirely within the CybORG/CAGE-4 simulator.
No real networks, credentials, or exploit code are involved.
Red-process actions are abstract primitives with no operational meaning outside
the simulator.
The anonymised replication package contains no offensive capability.



\section{Theoretical Foundations}
\label{sec:foundations}

\subsection{Markov Decision Process}

\begin{definition}[Markov Decision Process (MDP)]
A \emph{Markov Decision Process} (MDP) is a tuple
$(\mathcal{S},\mathcal{A},P,R,\gamma,\rho_0)$
where $\mathcal{S}$ is the state space, $\mathcal{A}$ the action space,
$P:\mathcal{S}\times\mathcal{A}\to\Delta(\mathcal{S})$ the transition kernel,
$R:\mathcal{S}\times\mathcal{A}\to\mathbb{R}$ the reward function,
$\gamma\in[0,1)$ the discount factor, and $\rho_0$ the initial state distribution.
\end{definition}

The discounted return from step $t$ is $G_t=\sum_{k=0}^{T-1-t}\gamma^k r_{t+k}$.
The state-value and action-value functions under policy $\pi$ are:
\begin{align}
V^\pi(s) &= \mathbb{E}_\pi[G_t\mid s_t{=}s],
\label{eq:vfunc}\\
Q^\pi(s,a) &= \mathbb{E}_\pi[G_t\mid s_t{=}s,a_t{=}a].
\label{eq:qfunc}
\end{align}
The advantage function $A^\pi(s,a)=Q^\pi(s,a)-V^\pi(s)$ has zero expectation under $\pi$.

\subsection{Policy Gradient and Actor-Critic}

\begin{theorem}[Policy Gradient~\citep{sutton2018reinforcement}]
For any differentiable policy $\pi_\theta$,
\begin{equation}
\nabla_\theta J(\theta) = \mathbb{E}_{\pi_\theta}\!\left[
\sum_{t=0}^{T-1}\nabla_\theta\log\pi_\theta(a_t\mid s_t)\,A^{\pi_\theta}(s_t,a_t)
\right].
\label{eq:pg_theorem}
\end{equation}
\end{theorem}

The temporal-difference residual at step $t$ is:
\begin{equation}
\delta_t = r_t + \gamma(1-d_t)V_\phi(s_{t+1}) - V_\phi(s_t),
\label{eq:td_residual}
\end{equation}
where $d_t\in\{0,1\}$ indicates episode termination.

\subsection{Independent Advantage Actor--Critic}
\label{sec:ia2c_foundation}

Independent Advantage Actor--Critic (IA2C) is included as a reward-only
actor--critic baseline; it is the independent-agent version of Advantage
Actor--Critic (A2C)~\citep{mnih2016asynchronous}.  Each Blue agent
maintains its own actor--critic network and optimiser, with no weight sharing
and no centralised critic.  For agent $i$, the policy and value function are
conditioned only on the local observation:
\begin{equation}
\begin{aligned}
a_t^i&\sim\pi_{\theta_i}(\cdot\mid o_t^i),\\
V_{\phi_i}(o_t^i)&\in\mathbb{R}.
\end{aligned}
\label{eq:ia2c_policy_value}
\end{equation}
Here $t$ indexes the environment step and $i\in\{1,\ldots,N\}$ indexes a
Blue agent.  The local observation is $o_t^i$, the sampled discrete action is
$a_t^i$, $\pi_{\theta_i}$ is agent $i$'s actor with parameters $\theta_i$, and
$V_{\phi_i}$ is its scalar value critic with parameters $\phi_i$.
The dot in $\pi_{\theta_i}(\cdot\mid o_t^i)$ denotes the full categorical
distribution over the agent's valid actions conditioned on $o_t^i$.
The discounted return target used by the implementation is
\begin{equation}
\begin{aligned}
\hat R_t^i&=\sum_{\tau=t}^{T-1}\gamma^{\tau-t}r_\tau^i,\\
\gamma&=0.99,
\end{aligned}
\label{eq:ia2c_return}
\end{equation}
where $\tau$ is a summation index over future steps and $T$ is the episode
horizon; episode termination resets the recursion.  The baseline advantage is
\begin{equation}
\begin{aligned}
\hat A_t^{i,\mathrm{A2C}}
&=\hat R_t^i-V_{\phi_i}(o_t^i),\\
\tilde A_t^{i,\mathrm{A2C}}
&=\frac{\hat A_t^{i,\mathrm{A2C}}-\mu_A^i}{\sigma_A^i+\varepsilon}.
\end{aligned}
\label{eq:ia2c_advantage}
\end{equation}
where $\mu_A^i$ and $\sigma_A^i$ are the empirical mean and standard deviation
of agent $i$'s batch advantages, and $\varepsilon$ is a small positive constant
used only for numerical stability.
The per-agent objective is the vanilla actor--critic loss
\begin{equation}
\begin{aligned}
\mathcal{L}_{\mathrm{IA2C}}^i
=&-\mathbb{E}_t\!\left[
\log\pi_{\theta_i}(a_t^i\mid o_t^i)\,
\tilde A_t^{i,\mathrm{A2C}}
\right] \\
&+\frac{c_{\mathrm{vf}}}{2}\,
\mathbb{E}_t\!\left[
\bigl(V_{\phi_i}(o_t^i)-\hat R_t^i\bigr)^2
\right] \\
&-c_{\mathrm{ent}}\,
\mathbb{E}_t\!\left[
H\!\left(\pi_{\theta_i}(\cdot\mid o_t^i)\right)
\right],
\end{aligned}
\label{eq:ia2c_loss}
\end{equation}
Here $\mathbb{E}_t$ denotes the empirical average over rollout timesteps and
$H(\pi_{\theta_i})$ is the action-distribution entropy used to encourage
exploration.  We use $c_{\mathrm{vf}}=0.5$, $c_{\mathrm{ent}}=0.01$,
gradient clipping at $0.5$, and RMSprop learning rate $7\times10^{-4}$.
Unlike PPO and MAPPO, IA2C does not use a clipped policy ratio; it is included
to show how a simpler independent actor--critic behaves under the same CAGE-4
observation, action-mask, reward, and cost-logging protocol.

\subsection{Generalised Advantage Estimation}

The Generalised Advantage Estimation (GAE) advantage is computed with
$\gamma{=}0.99$ and $\lambda_{\mathrm{GAE}}{=}0.95$
via the backward recursion~\citep{schulman2016high}:
\begin{align}
\hat{A}_T &= 0,
\label{eq:gae_base}\\
\hat{A}_t &= \delta_t + \gamma\lambda_{\mathrm{GAE}}(1-d_t)\,\hat{A}_{t+1},
\quad t=T{-}1,\ldots,0,
\label{eq:gae_recurrence}
\end{align}
with bootstrapped return target $\hat{R}_t = \hat{A}_t + V_\phi(s_t)$.

\subsection{Proximal Policy Optimisation}

Proximal Policy Optimisation (PPO) updates the policy via a clipped surrogate
that prevents destructively large steps~\citep{schulman2017proximal}.
Define the importance-sampling ratio:
\begin{equation}
\rho_t(\theta) = \frac{\pi_\theta(a_t\mid s_t)}{\pi_{\theta_{\mathrm{old}}}(a_t\mid s_t)}.
\label{eq:ratio}
\end{equation}
The numerator is the current policy probability of the sampled action and the
denominator is the probability under the frozen behaviour policy
$\theta_{\mathrm{old}}$ that generated the rollout.

\paragraph{Policy loss.}
\begin{equation}
\begin{aligned}
\mathcal{L}^{\mathrm{CLIP}}(\theta)
&= -\mathbb{E}_t\!\left[
\min\!\left(q_t,\bar q_t\right)
\right],\\
q_t
&= \rho_t(\theta)\hat{A}_t,\\
\bar q_t
&= \mathrm{clip}\!\left(\rho_t(\theta),1{-}\epsilon,1{+}\epsilon\right)
\hat{A}_t,\qquad \epsilon=0.2.
\end{aligned}
\label{eq:ppo_clip}
\end{equation}
Here $q_t$ is the unclipped policy-gradient term and $\bar q_t$ is the same
term after clipping the likelihood ratio to the interval
$[1-\epsilon,1+\epsilon]$.

\paragraph{Clipped value loss.}
To prevent catastrophic value-function updates:
\begin{equation}
\begin{aligned}
V_{\phi,\mathrm{clip}}
&= V_{\phi_{\mathrm{old}}}(s_t) + \Delta V_t,\\
\Delta V_t
&= \mathrm{clip}\!\left(
V_\phi(s_t)-V_{\phi_{\mathrm{old}}}(s_t),
-\epsilon,\epsilon
\right),
\end{aligned}
\label{eq:vclip}
\end{equation}
\begin{equation}
\begin{aligned}
\mathcal{L}^{\mathrm{VF}}(\phi)
&= \tfrac{1}{2}\,\mathbb{E}_t\!\left[
\max\!\left(e_t^2,\bar e_t^2\right)
\right],\\
e_t &= V_\phi(s_t)-\hat{R}_t,\\
\bar e_t &= V_{\phi,\mathrm{clip}}-\hat{R}_t .
\end{aligned}
\label{eq:value_loss}
\end{equation}

\paragraph{Total PPO objective.}
\begin{equation}
\mathcal{L}(\theta,\phi)
= \mathcal{L}^{\mathrm{CLIP}}(\theta)
+ c_{\mathrm{vf}}\,\mathcal{L}^{\mathrm{VF}}(\phi)
- c_{\mathrm{ent}}\,H[\pi_\theta],
\label{eq:ppo_total}
\end{equation}
where $c_{\mathrm{vf}}$ weights the critic loss and $c_{\mathrm{ent}}$ weights
the entropy bonus.
We use $c_{\mathrm{vf}}{=}0.5$, $c_{\mathrm{ent}}{=}0.005$ for ACD$^3$-GAT
and $c_{\mathrm{ent}}{=}0.01$ for the MAPPO variants.
Parameters are updated with Adam ($\eta{=}3\times10^{-4}$),
gradient norm clipped at $0.5$, over $4$ optimisation epochs.  MAPPO-family
baselines use mini-batch size $64$; the reported ACD$^3$-GAT optimisation
applies the same PPO objective over its concatenated episode batch as specified in
Section~\ref{sec:composite}.

\subsection{Multi-Agent Proximal Policy Optimisation (MAPPO) with Centralised Critic}

Multi-Agent Proximal Policy Optimisation (MAPPO) adopts Centralised Training
with Decentralised Execution (CTDE)~\citep{yu2022surprising}, following the
centralised-critic tradition in multi-agent actor-critic
methods~\citep{lowe2017multi}.
A centralised value function $V_\phi(\mathbf{o}_t)$ has access to all $N{=}5$
agents' observations $\mathbf{o}_t=(o_t^1,\ldots,o_t^N)$ during training,
while each actor $\pi_{\theta_i}$ operates on local $o_t^i$ at execution.
The centralised value function is:
\begin{equation}
\begin{aligned}
V_\phi(\mathbf{o}_t)
&= f_\phi(\mathbf{z}_t),\\
\mathbf{z}_t
&= \bigoplus_{i=1}^{5}\mathrm{Enc}^i_\phi(o_t^i)
\in\mathbb{R}^{320}.
\end{aligned}
\label{eq:mappo_critic}
\end{equation}
Here $\bigoplus$ denotes concatenation of five 64-dimensional per-agent
embeddings into the fused critic input $\mathbf{z}_t$, and
$f_\phi:\mathbb{R}^{320}\to\mathbb{R}$ is a two-layer multi-layer perceptron (MLP).
This separates training and execution information: the critic sees the
five-agent fused observation during optimisation, whereas each Blue actor
samples from its own local observation or context and the valid CybORG action
mask at execution time.

\subsection{Constrained Markov Decision Process}

\begin{definition}[Constrained Markov Decision Process (CMDP)~\citep{altman1999constrained}]
A Constrained Markov Decision Process (CMDP) augments an MDP with $K$ cost functions
$c^k:\mathcal{S}\times\mathcal{A}\to\mathbb{R}_{\geq 0}$ and budget
constraints $B_k>0$.
The feasible policy set is
$\Pi_\mathcal{C}=\{\pi: J_{c_k}(\pi)\leq B_k,\,\forall k\}$,
where $J_{c_k}(\pi)=\mathbb{E}_\pi[\sum_t\gamma^t c_t^k]$.
\end{definition}

\begin{proposition}[Strong duality~\citep{altman1999constrained}]
Under the linear programming relaxation of the discounted CMDP,
\begin{equation}
\max_\pi\min_{\boldsymbol\lambda\geq\mathbf{0}}\mathcal{L}(\pi,\boldsymbol\lambda)
=\min_{\boldsymbol\lambda\geq\mathbf{0}}\max_\pi\mathcal{L}(\pi,\boldsymbol\lambda),
\label{eq:strong_duality}
\end{equation}
where $\mathcal{L}(\pi,\boldsymbol\lambda)
=J_r(\pi)-\sum_k\lambda_k(J_{c_k}(\pi)-B_k)$.
\end{proposition}
This proposition motivates the Lagrangian update used in the algorithm, but
the implemented neural MARL problem is non-convex and partially observed.
Accordingly, the paper reports empirical episode-level violation rates rather
than claiming a formal strong-duality guarantee for ACD\textsuperscript{3}-GAT.

\subsection{Graph Neural Networks}

\paragraph{GraphSAGE.}
The MAPPO-GNN baseline uses the same observation-to-graph parser as the GAT
models, but replaces attention with a two-layer GraphSAGE encoder.  For layer
$\ell$, GraphSAGE aggregates neighbouring embeddings and applies a learned
linear map:
\begin{equation}
\mathbf{h}_u^{(\ell+1)}
=\mathrm{LN}\!\left(
\mathrm{ReLU}\!\left(
W_{\mathrm{sage}}^{(\ell)}
\bigl[\mathbf{h}_u^{(\ell)}\;\|\;
\mathrm{mean}_{v\in\mathcal{N}(u)}\mathbf{h}_v^{(\ell)}\bigr]
\right)\right).
\label{eq:graphsage_update}
\end{equation}
Here $u$ is the node being updated, $v$ indexes neighbouring nodes in
$\mathcal{N}(u)$, $\|$ denotes vector concatenation, and
$W_{\mathrm{sage}}^{(\ell)}$ is the learned linear map at layer $\ell$.
LN denotes layer normalisation and ReLU denotes the rectified linear unit.
The implementation uses hidden dimension $64$ in the first layer, output
dimension $64$ in the second layer, and global mean pooling over nodes to
obtain the per-agent embedding.

\paragraph{Graph Attention Networks.}
Graph neural networks can be understood as neural message-passing
architectures~\citep{gilmer2017neural}; GraphSAGE provides an inductive
neighbourhood-aggregation baseline~\citep{hamilton2017inductive}.
A Graph Attention Network (GAT) layer computes attention-weighted
neighbourhood aggregation~\citep{velickovic2018graph}.
Given node features $\{\mathbf{h}_u^{(\ell)}\}$,
the un-normalised attention score from $v$ to $u$ is:
\begin{equation}
e_{uv}^{(\ell)} = \mathrm{LeakyReLU}_{0.2}\!\left(
\mathbf{a}^{(\ell)\top}\!\left[
W^{(\ell)}\mathbf{h}_u^{(\ell)}\;\|\;W^{(\ell)}\mathbf{h}_v^{(\ell)}
\right]\right),
\label{eq:gat_score}
\end{equation}
Here $u$ is the destination node, $v$ is a neighbour contributing a message,
$W^{(\ell)}$ is the layer-$\ell$ feature projection, and
$\mathbf{a}^{(\ell)}$ is the learned attention vector.
The scores are normalised via softmax over $\mathcal{N}(u)$:
\begin{equation}
\alpha_{uv}^{(\ell)} = \frac{\exp(e_{uv}^{(\ell)})}
{\sum_{w\in\mathcal{N}(u)}\exp(e_{uw}^{(\ell)})}.
\label{eq:gat_attn}
\end{equation}
The coefficient $\alpha_{uv}^{(\ell)}$ is therefore the normalised attention
weight assigned to neighbour $v$ when updating node $u$.
The updated representation with $K_\ell$ attention heads ($\|$ = concatenation):
\begin{equation}
\mathbf{h}_u^{(\ell+1)} = \mathrm{LN}\!\left(\mathrm{ELU}\!\left(
\underset{m=1}{\overset{K_\ell}{\|}}
\sum_{v\in\mathcal{N}(u)}\alpha_{uv,m}^{(\ell)}W_m^{(\ell)}\mathbf{h}_v^{(\ell)}
\right)\right),
\label{eq:gat_update}
\end{equation}
where $m$ indexes the attention head, $W_m^{(\ell)}$ is the head-specific
projection, and ELU denotes the exponential linear unit.
In the implementation, the PyTorch Geometric GAT layers use their default
self-loop augmentation, so $\mathcal{N}(u)$ includes the node itself during
message passing.

Our encoder uses two layers:
Layer~1 with $K_1{=}4$ heads and per-head dimension $16$
(output $\in\mathbb{R}^{64}$);
Layer~2 with $K_2{=}1$ head and dimension $64$
(output $\in\mathbb{R}^{64}$).
The global graph embedding is obtained by mean-pooling:
$\mathbf{g} = \frac{1}{|V|}\sum_{v\in V}\mathbf{h}_v^{(2)}\in\mathbb{R}^{64}$.

\paragraph{Gated Recurrent Unit.}
The Gated Recurrent Unit (GRU) cell maps input
$\mathbf{x}_t\in\mathbb{R}^{d_x}$ and state
$\mathbf{h}_{t-1}\in\mathbb{R}^{d_h}$ as follows
\citep{cho2014learning}:
\begin{align}
\mathbf{z}_t &= \sigma(W_z\mathbf{x}_t + U_z\mathbf{h}_{t-1}),
\nonumber\\
\mathbf{r}_t &= \sigma(W_r\mathbf{x}_t + U_r\mathbf{h}_{t-1}),
\label{eq:gru_gates}\\
\tilde{\mathbf{h}}_t &= \tanh(W_h\mathbf{x}_t + U_h(\mathbf{r}_t\odot\mathbf{h}_{t-1})),
&&\label{eq:gru_candidate}\\
\mathbf{h}_t &= (1-\mathbf{z}_t)\odot\mathbf{h}_{t-1} + \mathbf{z}_t\odot\tilde{\mathbf{h}}_t.
\label{eq:gru_output}
\end{align}
Here $\mathbf{z}_t$ is the update gate, $\mathbf{r}_t$ the reset gate,
$\tilde{\mathbf{h}}_t$ the candidate state, $\sigma$ the logistic sigmoid,
and $\odot$ element-wise multiplication.
We use $d_x{=}64$ (graph embedding dimension) and $d_h{=}32$
(opponent latent dimension).



\section{Problem Formulation}
\label{sec:problem}

\subsection{CAGE-4 as a Constrained Decentralized Partially Observable Markov Decision Process (Dec-POMDP)}

We instantiate the Dec-POMDP formalism~\citep{oliehoek2016concise}
(Section~\ref{sec:foundations}) with
$N{=}5$ CAGE ``Blue'' agents, episode length $T{=}500$, and $\gamma{=}0.99$.
The global simulator state is not observed directly by the response policies.
At each step, each Blue agent receives a local binary observation, selects a
discrete response action, and the joint Blue action is applied together with
the CAGE Red transition dynamics in CybORG.
The learning problem is therefore to optimise decentralised Blue policies
from partial observations while satisfying episode-level operational budgets.
The MDP notation in Section~\ref{sec:foundations} is used as background:
the hidden simulator state $s_t$ defines the transition process, but the
implemented policies condition on local observations $o_t^i$ or on the
derived ACD\textsuperscript{3} context vector.  Thus equations written in
state notation should be read as their partially observed implementation
counterparts after replacing $s_t$ by $o_t^i$, $\mathbf{o}_t$, or
$\mathrm{ctx}_t^i$ as specified below.

\paragraph{Notation used throughout.}
Subscript $t$ indexes an environment step, superscript $i$ indexes a Blue
agent, superscript $k$ indexes an operational constraint, and $e$ indexes an
episode.
The policy $\pi_\theta$ is executed independently by each Blue agent.  The
formal objective is written in terms of mean team reward and episode-level
operational budgets, while the implemented ACD$^3$ PPO update stores
per-agent reward and cost streams and then reports the same costs after
aggregating over agents and time.
Rewards are discounted when constructing PPO advantages; operational costs are
recorded as undiscounted totals because the budgets are governance limits
rather than reward-shaping terms.
Thus $r_t^i$ denotes the simulator reward observed by agent $i$,
$\bar r_t=\frac{1}{5}\sum_i r_t^i$ is the team reward used in the objective,
$c_{t}^{i,k}$ is the immediate cost proxy for agent $i$, and
$c_t^k=\sum_i c_t^{i,k}$ denotes the audited team cost at step $t$.
The episode total $\sum_t c_t^k$ is compared with budget $B_k$.
This distinction is important for reproducibility: the optimiser uses
discounted reward advantages, while the reported safety metrics and Lagrange
multiplier updates use undiscounted episode cost totals.

\paragraph{Observation space.}
All agents share a uniform observation dimension $d{=}210$.
The observation $o_t^i\in\{0,1\}^{210}$ has the structure:
\begin{equation}
\begin{aligned}
o_t^i
&=\bigl[\phi_t\;\|\;B_0\;\|\;B_1\;\|\;B_2\;\|\;M\bigr],\\
&\quad
\phi_t\in\{0,1,2\},\quad B_j\in\{0,1\}^{59},\quad
M\in\{0,1\}^{32}.
\end{aligned}
\label{eq:obs_structure}
\end{equation}
Here $\phi_t$ is the mission phase (Preplanning, MissionA, MissionB),
$B_j$ is subnet block $j$, and $M$ encodes inter-agent messages
($4$ agents $\times$ $8$ bits).

Agents~$0$--$3$ each monitor one subnet (blocks $B_1$, $B_2$ are zero-padded);
agent~$4$ monitors three subnets (all three blocks are real).
Each $59$-bit block encodes:
\begin{equation}
\begin{aligned}
B_j
&=\bigl[s_j\;\|\;b_j\;\|\;m_j\;\|\;p_j\;\|\;n_j\bigr],\\
s_j,b_j,m_j&\in\{0,1\}^{9},\qquad
p_j,n_j\in\{0,1\}^{16}.
\end{aligned}
\label{eq:block_structure}
\end{equation}
Here $s_j$ is the subnet one-hot vector, $b_j$ marks blocked subnets,
$m_j$ stores communications-policy mismatch, and $p_j[h]$ (resp.\ $n_j[h]$)
equals $1$ if a malicious process (resp.\ network connection) event has been
detected on host $h$ in subnet $j$.
Hosts are ordered: server hosts $h\in\{0,\ldots,5\}$ followed by user
hosts $h\in\{6,\ldots,15\}$.
This is the observation-to-graph interface used throughout the paper:
the flat 210-bit observation is environment-specific, while the resulting
host--subnet graph representation is the reusable input expected by the graph
encoders and Graph Counterfactual Risk Propagation (G-CRP) screen.

\paragraph{Action space.}
Each agent selects from $|\mathcal{A}^i|\leq 242$ discrete actions.
Actions decompose into $K{=}9$ types:
\begin{equation}
\begin{aligned}
\mathcal{T}=\{&
\textsc{Sleep},\;\textsc{Monitor},\;\textsc{Analyse},\\
&\textsc{Remove},\;\textsc{Restore},\;\textsc{BlockZone},\\
&\textsc{AllowZone},\;\textsc{DeployDecoy},
\textsc{Other}\}.
\end{aligned}
\label{eq:action_types}
\end{equation}
For compact notation, \textsc{BlockZone} and \textsc{AllowZone} denote the
simulator actions \textsc{BlockTrafficZone} and \textsc{AllowTrafficZone}.
The time durations are $\delta:\mathcal{T}\to\{1,2,3,5\}$:
Sleep, Monitor, BlockZone, and AllowZone take one simulator
step; Analyse and DeployDecoy take two; Remove takes three; Restore takes
five.

\subsection{Operational Cost Signals}
\label{sec:costs}

\begin{definition}[Three operational cost signals]
At step $t$ for agent $i$, with action $a_t^i$ and observation $o_t^i$:
\begin{align}
c_{t}^{i,\mathrm{down}} &= \mathbf{1}[\tau(a_t^i)=\mathrm{Restore}],
\label{eq:c_down}\\
c_{t}^{i,\mathrm{fw}} &=
\mathbf{1}[\tau(a_t^i)\in\{\mathrm{BlockZone},\mathrm{AllowZone}\}],
\label{eq:c_fw}\\
c_{t}^{i,\mathrm{fp}} &= c_{t}^{i,\mathrm{down}}\cdot
\mathbf{1}\!\left[\sum_{j,h}(p_j[h]\vee n_j[h])=0\right],
\label{eq:c_fp}
\end{align}
where $\tau(a)$ denotes the action type of $a$.
The indicator in~\eqref{eq:c_fp} is zero when at least one malicious flag is
active, so a Restore action is counted as false-positive only when no alert
evidence is visible in the decoded observation.
The team cost audited in the benchmark is
$c_t^k=\sum_i c_t^{i,k}$.
These are proxy labels derived from simulator action names and visible alerts:
\textsc{Restore} contributes downtime, \textsc{Restore} without visible alert
evidence contributes false-positive response cost, and
\textsc{BlockTrafficZone} and \textsc{AllowTrafficZone} contribute firewall-change
cost.  A violation of any one budget is a violation of the episode safety
contract.
\end{definition}

\begin{definition}[Safety contract]
Policy $\pi$ satisfies the safety contract in episode $e$ iff
\begin{equation}
\sum_{t=0}^{T-1} c_t^k\leq B_k,\quad\forall k\in\mathcal{K},
\label{eq:safety_contract}
\end{equation}
with $\mathcal{K}=\{\mathrm{down},\mathrm{fw},\mathrm{fp}\}$,
$B_{\mathrm{down}}{=}50$, $B_{\mathrm{fw}}{=}20$, $B_{\mathrm{fp}}{=}10$.
The per-constraint violation rate is $P(\mathrm{viol}^k)
=\mathbb{P}(\sum_t c_t^k>B_k)$.
\end{definition}

\subsection{Constrained Joint Objective}

\begin{equation}
\begin{aligned}
\max_\pi\;J_r(\pi)
&=\mathbb{E}_\pi\!\left[
\sum_{t=0}^{T-1}\gamma^t\bar{r}_t
\right],\\
\text{s.t.}\quad
J_{c_k}(\pi)
&=\mathbb{E}_\pi\!\left[
\sum_{t=0}^{T-1}c_t^k
\right]\leq B_k,\quad\forall k .
\end{aligned}
\label{eq:constrained_obj}
\end{equation}
where $\bar{r}_t=\frac{1}{5}\sum_{i=1}^{5}r_t^i$ is the mean team reward.
Operational costs are audited as undiscounted episode totals, matching the
budget accounting used by the shield and evaluation metrics.

\paragraph{Lagrangian relaxation following the constrained objective.}
\begin{equation}
\begin{aligned}
\mathcal{L}(\pi,\boldsymbol\lambda)
&=J_r(\pi)
-\sum_{k\in\mathcal{K}}\lambda_k
\bigl(J_{c_k}(\pi)-B_k\bigr),\\
\boldsymbol\lambda&\geq\mathbf{0}.
\end{aligned}
\label{eq:lagrangian}
\end{equation}

\paragraph{Dual gradient ascent.}
Multipliers are updated after each episode batch:
\begin{equation}
\begin{aligned}
\lambda_k
&\leftarrow
\bigl[\lambda_k+\eta_\lambda(\hat{J}_{c_k}-B_k)\bigr]_+,
\qquad \eta_\lambda=0.01,\\
\hat{J}_{c_k}
&=\frac{1}{M}\sum_{e=1}^{M}\sum_{t=0}^{T_e-1}c_{e,t}^k .
\end{aligned}
\label{eq:lambda_update}
\end{equation}
where $M$ is the number of complete episodes in the PPO update batch.

\subsection{Graph Observation Encoding}
\label{sec:graph_encoding}

\paragraph{Graph construction.}
For each agent $i$ and step $t$, we build a graph
$G_t^i=(V,E,X)$ from $o_t^i$:
\begin{itemize}
\item\textbf{Nodes:} One node per host slot per monitored subnet
  ($16$ slots: $6$ server, $10$ user), plus one subnet-level node.
  Agents~$0$--$3$: $|V|{=}17$. Agent~$4$: $|V|{=}51$.
\item\textbf{Edges:} Bidirectional star between each host node and its
  subnet node; inter-subnet edges for agent~$4$ where the communication
  policy permits.
\item\textbf{Node features:} $\mathbf{x}_v\in\mathbb{R}^9$ defined below.
\end{itemize}

\paragraph{Node feature vector.}
For node $v$ corresponding to host index $h$ in subnet $j$:
\begin{equation}
\begin{aligned}
\mathbf{x}_v
&=\bigl[
x_v^{\mathrm{srv}},x_v^{\mathrm{usr}},x_v^{\mathrm{sub}},
p_j[h],\\
&\qquad n_j[h],\bar{m}_j,\mathbf{e}_{\phi_t}^{\top}
\bigr]^\top\in\mathbb{R}^9 ,
\end{aligned}
\label{eq:node_features}
\end{equation}
where $x_v^{\mathrm{srv}}{=}1$ iff $h\in\{0,\ldots,5\}$,
$x_v^{\mathrm{usr}}{=}1$ iff $h\in\{6,\ldots,15\}$,
$x_v^{\mathrm{sub}}{=}1$ for the subnet summary node,
$\bar{m}_j$ is the mean communications-policy mismatch in subnet $j$,
and $\mathbf{e}_{\phi_t}\in\{0,1\}^3$ is the mission-phase one-hot vector.

\paragraph{GAT encoder.}
Applying Equations~\eqref{eq:gat_score}--\eqref{eq:gat_update} with
$d_{\mathrm{in}}{=}9$, $K_1{=}4$ heads, per-head dimension $16$
(Layer~1 output: $\mathbb{R}^{64}$),
then $K_2{=}1$ head, dimension $64$
(Layer~2 output: $\mathbb{R}^{64}$):
\begin{equation}
\mathbf{g}_t^i = \frac{1}{|V|}\sum_{v\in V}\mathbf{h}_v^{(2)}
\in\mathbb{R}^{64}.
\label{eq:global_emb}
\end{equation}

\subsection{Factorised Action Policy}
\label{sec:factorized}

Every action $a\in\{0,\ldots,|\mathcal{A}^i|-1\}$ decomposes into
a type $\tau(a)\in\mathcal{T}$ and a target index
$\nu(a)\in\{0,\ldots,M_{\tau(a)}-1\}$,
where $M_\tau$ is the number of valid targets for type $\tau$
and $M_{\max}=\max_\tau M_\tau$.

The factorised policy is:
\begin{align}
\pi_\theta(a\mid\mathrm{ctx}_t)
&=\pi_\theta^{\mathrm{type}}(\tau(a)\mid\mathrm{ctx}_t)\cdot
\pi_\theta^{\mathrm{tgt}}(\nu(a)\mid\mathrm{ctx}_t,\tau(a)),
\label{eq:factorized_policy}\\
\pi_\theta^{\mathrm{type}}(\tau\mid\mathrm{ctx})
&=\frac{\exp(w_\tau^\top\mathrm{ctx}+\alpha_\tau)\cdot\mathbf{1}_\tau}
{\sum_{\tau'}\exp(w_{\tau'}^\top\mathrm{ctx}+\alpha_{\tau'})\cdot\mathbf{1}_{\tau'}},
\label{eq:type_head}\\
\pi_\theta^{\mathrm{tgt}}(\nu\mid\mathrm{ctx},\tau)
&=\frac{\exp(f_\nu(\mathrm{ctx},\mathbf{e}_\tau)_\nu)\cdot\mathbf{1}_\nu}
{\sum_{\nu'}\exp(f_\nu(\mathrm{ctx},\mathbf{e}_\tau)_{\nu'})\cdot\mathbf{1}_{\nu'}},
\label{eq:target_head}
\end{align}
where $\mathbf{1}_\tau,\mathbf{1}_\nu$ mask invalid types/targets,
$\mathbf{e}_\tau\in\mathbb{R}^{16}$ is a learnable type embedding,
and $f_\nu:\mathbb{R}^{120+16}\to\mathbb{R}^{M_{\max}}$ is a two-layer MLP.

The joint log-probability is
$\log\pi_\theta(a\mid\mathrm{ctx})
=\log\pi_\theta^{\mathrm{type}}(\tau(a))+\log\pi_\theta^{\mathrm{tgt}}(\nu(a))$,
and the entropy decomposes as
$H[\pi_\theta]=H[\pi^{\mathrm{type}}]+\mathbb{E}_\tau[H[\pi^{\mathrm{tgt}}(\cdot\mid\tau)]]$.



\section{ACD$^3$-GAT}
\label{sec:method}

\subsection{Method Overview}

ACD$^3$-GAT solves the constrained Decentralized Partially Observable Markov
Decision Process (Dec-POMDP) in
Equation~\eqref{eq:constrained_obj} by combining four mechanisms.
First, each agent converts its binary CAGE-4 observation into a host--subnet
graph and encodes it with a Graph Attention Network (GAT).
Second, the graph embedding is concatenated with a recurrent opponent
embedding, the remaining budget state, and an uncertainty signal to form the
policy context.
Third, a factorised actor proposes an action type and target, while a set of
reward, cost, tail-risk, and exploitability critics supplies PPO training
signals.
Fourth, before execution, a graph-risk shield evaluates admissible
counterfactual actions with $Q_{\mathrm{safe}}$ and replaces unsafe proposals
when the safety contract requires it.

\begin{figure*}[!t]
\centering
\includegraphics[width=0.98\textwidth]{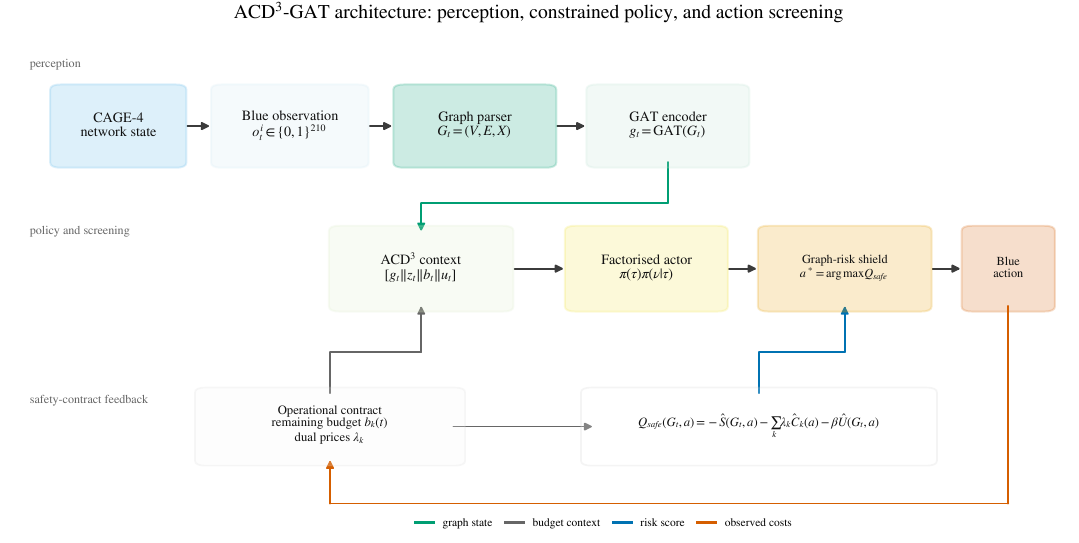}
\caption{ACD\textsuperscript{3}-GAT system architecture.
  Partial CAGE-4 observations are parsed into host--subnet graphs, encoded
  with graph attention, combined with opponent, budget, and uncertainty
  context, and passed to a factorised actor.  The graph-risk shield screens
  the proposed action with current safety-contract multipliers; observed
  costs update the remaining budgets and the next dual step.  The parser and
  action layout are simulator-specific interfaces, while the graph encoder,
  constrained policy, and safety-contract feedback are reusable components.}
\label{fig:system_overview}
\end{figure*}

\begin{figure*}[!t]
\centering
\includegraphics[width=0.98\textwidth]{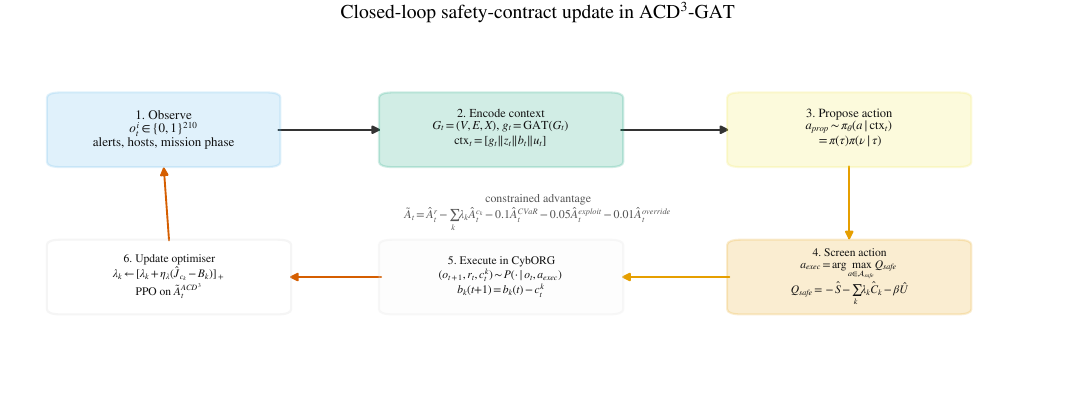}
\caption{Closed-loop safety-contract update used by ACD\textsuperscript{3}-GAT.
  Each decision observes the cyber range, encodes graph and budget context,
  proposes and screens an action, executes the selected response in CybORG,
  and stores the resulting reward, costs, and context.  Remaining budgets
  update at the next step, while Lagrange multipliers update after the PPO
  episode batch, separating execution-time safety from slower policy learning.}
\label{fig:acd3_control_loop}
\end{figure*}

Figures~\ref{fig:system_overview} and~\ref{fig:acd3_control_loop} give the
architectural and closed-loop views used during training and evaluation.
Figure~\ref{fig:system_overview} should be read from left to right as the
implemented data path.
The CAGE-4 network state first becomes the binary Blue observation $o_t^i$;
the parser converts that simulator-specific vector into a host--subnet graph
$G_t=(V,E,X)$. The GAT encoder produces the graph embedding
$\mathbf{g}_t$, and the policy context concatenates
$\mathbf{g}_t$ with recurrent opponent state $\mathbf{z}_t$, budget state
$\mathbf{b}_t$, and uncertainty $\mathbf{u}_t$ before the factorised actor
samples an action type and target.
The lower path in the same figure is the safety path: the shield scores
candidate actions with $Q_{\mathrm{safe}}$, using graph risk, operational
cost multipliers, and remaining budgets, and only then emits the action sent
to CybORG.
The feedback arrows therefore correspond exactly to the two quantities that
make ACD\textsuperscript{3}-GAT different from reward-only graph MARL:
executed-action costs update the remaining episode budgets at the next
decision step, while batch-level costs update the Lagrange multipliers used by
the next PPO optimisation.

Figure~\ref{fig:acd3_control_loop} translates the same architecture into the
temporal order of one interaction step:
observe $o_t^i$, parse $G_t^i$, encode $\mathbf{g}_t^i$, build
$\mathrm{ctx}_t^i$, propose $a_{\mathrm{prop}}$, screen it into
$a_{\mathrm{exec}}$, execute in CybORG, store reward and costs, and update
budgets and multipliers on their respective time scales.
The environment produces observations, as the policy proposes a response action, the shield selects the executable action, the CybORG returns reward and operational costs, and the remaining budgets and Lagrange multipliers determine how the next decision is evaluated. The
Algorithm~\ref{alg:acd3} gives the same procedure in executable order.
The equations below define each quantity used in that loop.  Together with the problem notation in Section~\ref{sec:problem}, they specify the observation parser, graph encoder, recurrent context, actor factorisation, critic targets, composite advantage, action shield, and dual update used by the reported implementation.
The figure is deliberately architectural rather than decorative: learned
modules are the GAT, recurrent context, actor, and critic heads; deterministic or governance modules are the observation parser, budget accounting, and Graph Counterfactual Risk Propagation (G-CRP) shield.  The graph parser and valid-action layout are environment interfaces, while the constrained policy, graph encoder, cost critics, and
shielding equations are the reusable ACD\textsuperscript{3} method components.


\subsection{From Safety Contract to Policy Update}
\label{sec:method_derivation}

The implemented update follows directly from the constrained objective in
Equation~\eqref{eq:constrained_obj}.
For a fixed multiplier vector $\boldsymbol{\lambda}$, maximising the
Lagrangian in Equation~\eqref{eq:lagrangian} is equivalent to maximising a
reward signal in which operational costs reduce the policy advantage.
Using the policy-gradient identity in Equation~\eqref{eq:pg_theorem}, the
unconstrained reward advantage $\hat A_t^r$ is therefore replaced by a
Lagrangian advantage
\begin{equation}
\hat A_t^{\mathrm{Lag}}
=\hat A_t^r-\sum_{k\in\{\mathrm{down},\mathrm{fw},\mathrm{fp}\}}
\lambda_k\hat A_t^{c_k},
\label{eq:lag_advantage_method}
\end{equation}
where $\hat A_t^r$ and $\hat A_t^{c_k}$ are both computed with the same GAE
recursion but from different scalar streams: simulator reward for
$\hat A_t^r$ and operational cost proxy $c_t^{i,k}$ for $\hat A_t^{c_k}$.
This is the first point where the safety contract enters the policy update:
actions that improve security reward can still be discouraged when they
increase expected downtime, firewall disruption, or false-positive Restore
cost.

ACD\textsuperscript{3}-GAT then augments this Lagrangian advantage with three
additional signals that are present in the implementation.
CVaR reweighting emphasises the worst-return episodes in the PPO batch,
adaptive opposing-policy evaluation can contribute an exploitability signal, and the
override term penalises excessive shield or human-governance burden.
The resulting scalar used in the PPO surrogate is
\begin{equation}
\begin{aligned}
\hat A_t^{\mathrm{ACD}^{3}}
&= \hat A_t^{\mathrm{Lag}}
   - \beta \hat A_t^{\mathrm{CVaR}} \\
&\quad
   - \eta \hat A_t^{\mathrm{exploit}}
   - \mu \hat A_t^{\mathrm{override}},\\[-0.25ex]
(\beta,\eta,\mu) &= (0.1,0.05,0.01).
\end{aligned}
\label{eq:acd3_from_lag}
\end{equation}
Before entering the clipped PPO objective, this advantage is normalised within
the batch:
\begin{equation}
\begin{aligned}
\tilde A_t
&=\frac{\hat A_t^{\mathrm{ACD}^3}-\mu_A}{\sigma_A+\varepsilon},\\
\varepsilon&=10^{-8}.
\end{aligned}
\label{eq:acd3_norm}
\end{equation}
Equation~\eqref{eq:ppo_clip} is then applied with $\tilde A_t$ in place of
the standard reward-only advantage.

The multiplier update closes the loop.
After a PPO batch, the observed episode cost estimate $\hat J_{c_k}$ is
compared with its budget $B_k$:
\begin{equation}
\begin{aligned}
\lambda_k
&\leftarrow
\bigl[\lambda_k+\eta_\lambda(\hat J_{c_k}-B_k)\bigr]_+,\\
\eta_\lambda&=0.01 .
\end{aligned}
\label{eq:acd3_dual_loop}
\end{equation}
If a cost remains below budget, the corresponding multiplier is unchanged or
decreases toward zero through projection; if it exceeds budget, the multiplier
increases and future PPO updates penalise that cost more strongly through
Equation~\eqref{eq:lag_advantage_method}.
In parallel, the action shield in Section~\ref{sec:shield} applies the current
remaining budgets $b_k(t)$ at decision time.
Thus the method has two safety mechanisms with different time scales:
Lagrangian learning shapes the policy across batches, while the shield prevents
budget-exhausting actions during an episode.

\subsection{Context Vector}
\label{sec:context}

The ACD$^3$-GAT policy conditions on a $120$-dimensional context vector
assembling four representations:
\begin{equation}
\mathrm{ctx}_t^i
=\bigl[\mathbf{g}_t^i\;\|\;\mathbf{z}_t^i\;\|\;\mathbf{b}_t\;\|\;\mathbf{u}_t\bigr]
\in\mathbb{R}^{64+32+16+8},
\label{eq:ctx}
\end{equation}
where each component is derived from $o_t^i$, the recurrent episode history,
or the current safety-contract state.

\paragraph{Graph embedding $\mathbf{g}_t^i\in\mathbb{R}^{64}$.}
Obtained by the two-layer GAT encoder
(Equations~\ref{eq:gat_score}--\ref{eq:global_emb})
applied to $G_t^i$.

\paragraph{Opponent embedding $\mathbf{z}_t^i\in\mathbb{R}^{32}$.}
A GRU (Equations~\ref{eq:gru_gates}--\ref{eq:gru_output}) encodes the
recent observation trajectory:
\begin{equation}
\mathbf{z}_t^i = \mathrm{GRU}_{64\to32}(\mathbf{g}_t^i,\,\mathbf{z}_{t-1}^i),
\quad \mathbf{z}_0^i=\mathbf{0}.
\label{eq:opponent_gru}
\end{equation}
This provides an implicit, online estimate of the CAGE Red process without
requiring explicit strategy identification.
When Red-process strategy labels $y\in\{0,\ldots,3\}$ and a non-zero auxiliary
weight are available (e.g., from league training), an auxiliary
classification loss
\begin{equation}
\mathcal{L}_{\mathrm{opp}}
=-\sum_{j=0}^{3}y_j\log\hat{y}_j(\mathbf{z}_t)
\label{eq:opp_aux}
\end{equation}
encourages $\mathbf{z}_t$ to encode discriminative opponent information.
Here $j$ indexes the Red-process class, $y_j$ is the one-hot target label,
and $\hat y_j(\mathbf{z}_t)$ is the classifier's predicted probability for
class $j$ from the recurrent state.

\paragraph{Budget embedding $\mathbf{b}_t\in\mathbb{R}^{16}$.}
The remaining operational budgets are embedded via a linear layer:
\begin{equation}
\begin{aligned}
\mathbf{b}_t
&=\tanh\!\left(W_b\,\mathbf{q}_t+\mathbf{d}_b\right),\\
\mathbf{q}_t
&=\left[
\tfrac{b_{\mathrm{down}}(t)}{B_{\mathrm{down}}},
\tfrac{b_{\mathrm{fw}}(t)}{B_{\mathrm{fw}}},
\tfrac{b_{\mathrm{fp}}(t)}{B_{\mathrm{fp}}}
\right]^\top,\\
W_b&\in\mathbb{R}^{16\times 3},
\qquad \mathbf{d}_b\in\mathbb{R}^{16}.
\end{aligned}
\label{eq:budget_emb}
\end{equation}
where $b_k(t)=\max(0,B_k-\sum_{\tau=0}^{t-1}c_\tau^k)$.
The network therefore receives normalised remaining budgets in $[0,1]$ rather
than raw episode totals.

\paragraph{Uncertainty signal $u_t\in[0,1]$ and embedding $\mathbf{u}_t\in\mathbb{R}^{8}$.}
The entropy of the action-type distribution conditioned on $\mathbf{g}_t$
alone (computed before the full context, avoiding circularity) is:
\begin{equation}
\begin{aligned}
u_t
&= \frac{
H\!\left(\mathrm{softmax}(W_u\mathbf{g}_t^i+\mathbf{d}_u)\right)
}{\log K},\\
W_u&\in\mathbb{R}^{9\times64},
\qquad \mathbf{d}_u\in\mathbb{R}^{9},
\qquad K=9 .
\end{aligned}
\label{eq:uncertainty}
\end{equation}
normalised to $[0,1]$.
A learned affine map
$\mathbf{u}_t=W_{\mathrm{ue}}\,u_t+\mathbf{d}_{\mathrm{ue}}\in\mathbb{R}^8$
embeds it into the context.
High $u_t$ indicates that the graph state alone is insufficient to
determine the action type; it is used as contextual evidence for the policy
and confidence monitor.
In the control loop, $u_t$ is therefore not a separate objective: it is a
compact uncertainty feature that enters the context vector and can trigger the
governance fallback when the configured confidence or G-CRP uncertainty
thresholds are exceeded.

\subsection{Multi-Objective Critic Set}
\label{sec:critics}

ACD$^3$-GAT maintains four distinct critic heads, all conditioned on
$\mathrm{ctx}_t$:
\begin{align}
V_r(\mathrm{ctx}_t)&\in\mathbb{R},
\label{eq:vr}\\
V_c^k(\mathrm{ctx}_t)&\in\mathbb{R},\quad k\in\{1,2,3\},
\label{eq:vc}\\
V_{\mathrm{CVaR}}(\mathrm{ctx}_t)&\in\mathbb{R},
\label{eq:vcvar}\\
V_{\mathrm{exploit}}(\mathrm{ctx}_t)&\in\mathbb{R}.
\label{eq:vexploit}
\end{align}
Each critic head is a two-layer MLP with hidden dimension $64$.
The reward head predicts the security-return baseline, the three cost
critics correspond to downtime, firewall-change cost, and false-positive
response cost, the CVaR head supports tail-risk accounting, and the
exploitability head stores the adaptive Red-process signal when enabled.
The cost critics are stacked as
$\mathbf{V}_c=(V_c^1,V_c^2,V_c^3)^\top\in\mathbb{R}^3$.

\subsection{Composite ACD$^3$ Advantage}
\label{sec:composite}

\paragraph{Per-objective GAE advantages.}
Using the recursion in Equations~\eqref{eq:gae_base}--\eqref{eq:gae_recurrence}:
\begin{align}
\hat{A}_t^r&=\mathrm{GAE}(\{r_\tau^i\},V_r;\gamma{=}0.99,\lambda_{\mathrm{GAE}}{=}0.95),
\label{eq:adv_r}\\
\hat{A}_t^{c_k}&=\mathrm{GAE}(\{c_\tau^{i,k}\},V_c^k;\gamma{=}0.99,\lambda_{\mathrm{GAE}}{=}0.95).
\label{eq:adv_c}
\end{align}
During ACD$^3$ optimisation these streams are computed per Blue agent from the
stored rollout of that agent; the reported episode metrics aggregate the same
cost proxies across agents and time.

\paragraph{CVaR advantage.}
In each batch of $M{=}8$ episodes ranked by return,
the $k^*{=}\max(1,\lfloor0.1M\rfloor)$ worst episodes receive
weight $w_e{=}1/k^*$; all others receive $w_e{=}0$:
\begin{equation}
\hat{A}_t^{\mathrm{CVaR}} = w_e\hat{A}_t^r.
\label{eq:adv_cvar}
\end{equation}
This is an episode-level reweighting of the policy-gradient signal rather than
a separate distributional value estimator.

\paragraph{Exploitability advantage.}
When adaptive opposing-policy evaluation is enabled, a short Red PPO mini-loop is run
every $I_{\mathrm{exploit}}$ episodes (default $50$;
3 updates $\times$ 2 episodes each):
\begin{equation}
\hat{A}_t^{\mathrm{exploit}}
=0.1\,\bar\delta,\quad
\bar\delta=U_{\mathrm{Blue}}(0)-U_{\mathrm{Blue}}(3),
\label{eq:adv_exploit}
\end{equation}
where $U_{\mathrm{Blue}}(k)$ is Blue's mean return after $k$ Red-process adaptation steps
and $\bar\delta$ is the running mean over the last $10$ stored measurements.
If no adaptive opposing-policy measurement has been stored, this term is zero.
In the reported experiments, coupled adaptive Red-process evaluation is used primarily as a
post-training stress test rather than as the headline optimisation target.

\paragraph{Override advantage.}
The centred override indicator penalises excessive analyst burden:
\begin{equation}
\hat{A}_t^{\mathrm{override}}
=\mathbf{1}[\text{shield triggered at }t]-\bar\omega,
\label{eq:adv_override}
\end{equation}
where $\bar\omega$ is the mean override rate in the current batch.

\paragraph{ACD$^3$ composite advantage.}
Substituting the terms above into Equation~\eqref{eq:acd3_from_lag} gives the
implemented scalar advantage:
\begin{equation}
\begin{aligned}
\hat{A}_t^{\mathrm{ACD}^3}
&=\hat{A}_t^r
-\sum_{k=1}^{3}\lambda_k\hat{A}_t^{c_k}
-0.1\,\hat{A}_t^{\mathrm{CVaR}}\\
&\quad
-0.05\,\hat{A}_t^{\mathrm{exploit}}
-0.01\,\hat{A}_t^{\mathrm{override}} .
\end{aligned}
\label{eq:acd3_advantage}
\end{equation}
Normalised before use:
\begin{equation}
\begin{aligned}
\tilde{A}_t
&=\frac{\hat{A}_t^{\mathrm{ACD}^3}-\mu_A}
        {\sigma_A+\varepsilon},\\
\varepsilon&=10^{-8}.
\end{aligned}
\label{eq:acd3_advantage_norm}
\end{equation}

\paragraph{PPO update.}
Equation~\eqref{eq:ppo_clip} is applied to $\tilde{A}_t$ with $\epsilon{=}0.2$,
$c_{\mathrm{vf}}{=}0.5$, $c_{\mathrm{ent}}{=}0.005$, and $4$ epochs over the
concatenated episode batch.  The same update also fits the reward, cost,
CVaR, and exploitability critic heads using clipped value losses.

\subsection{Budget-Aware Counterfactual Shield}
\label{sec:shield}

\paragraph{Remaining budget.}
The remaining budget is defined as:
\begin{equation}
b_k(t)=\max(0,B_k-\sum_{\tau=0}^{t-1}c_\tau^k)    
\end{equation}

\paragraph{Admissible action set.}
The reported shield uses the cost proxies as a hard exhaustion guard: once a
budget has been depleted, actions with positive immediate proxy cost for that
budget are blocked.  Costly actions before exhaustion remain admissible, but
are penalised by the Lagrangian and G-CRP score.
\begin{equation}
\begin{aligned}
\mathcal{A}_{\mathrm{safe}}(t)
&=\bigl\{a\in\mathcal{A}^i:\;
b_k(t)>0\\
&\lor\;\hat{c}_k(a)=0,\;\forall k
\bigr\},
\end{aligned}
\label{eq:safe_set}
\end{equation}
where $\hat{c}_k(a)$ is the immediate cost proxy
(Equations~\ref{eq:c_down}--\ref{eq:c_fp}).
In this expression, $\mathcal{A}^i$ is agent $i$'s discrete action set,
$b_k(t)$ is the remaining budget for constraint $k$ before action execution,
and the condition is applied for all three operational costs.

\paragraph{Shielded proposal.}
The policy first proposes an action from the factorised actor.  For efficiency,
the reported G-CRP shield ranks one representative candidate per action type.
It accepts the proposal when its action type is the top ranked admissible type;
otherwise it replaces the proposal with the highest-ranked admissible type, or
falls back to \textsc{Sleep} when no safe action is available:
\begin{equation}
\begin{aligned}
\pi_{\mathrm{prop}}(a\mid o_t^i)
&=\pi_\theta(a\mid\mathrm{ctx}_t),\\
a_{\mathrm{exec}}
&=\mathrm{Shield}\!\left(
a_{\mathrm{prop}},G_t,\mathcal{A}_{\mathrm{safe}}(t)
\right).
\end{aligned}
\label{eq:shielded_policy}
\end{equation}
Here $\pi_{\mathrm{prop}}$ is the stochastic policy distribution before
screening, $a_{\mathrm{prop}}$ is the sampled proposal, and
$a_{\mathrm{exec}}$ is the action actually submitted to CybORG after applying
the admissible set and G-CRP ranking.
When the shield replaces the proposed action, the implementation stores and
updates PPO on the executed action by recomputing its log-probability under
the current policy and action mask.  Thus the trajectory logger, cost
accounting, and likelihood ratio are all tied to the action actually submitted to CybORG. The shield is nevertheless a non-differentiable intervention and can bias the policy-gradient estimate, this is why override frequency is recorded and penalised in Equation~\eqref{eq:adv_override}. The rule blocks after the relevant remaining budget is exhausted, so episode-level compliance is evaluated empirically through the violation-rate metrics rather than asserted analytically from the shield alone.

\subsection{G-CRP: Graph Counterfactual Risk Propagation}
\label{sec:grp}

The rule shield~\eqref{eq:safe_set} reacts to cost proxies but cannot
estimate how the network's compromise state will evolve.
G-CRP provides predictive risk scoring via graph propagation.

\paragraph{Node risk beliefs.}
Given the proc and net flags in $o_t^i$:
\begin{equation}
\hat{p}_v(t)=\tfrac{1}{2}\,p_j[h(v)]+\tfrac{1}{2}\,n_j[h(v)]\in[0,1].
\label{eq:risk_belief}
\end{equation}
Here $h(v)$ maps host node $v$ to its decoded host slot, while $p_j[h]$ and
$n_j[h]$ are the process-alert and network-alert bits for host $h$ in subnet
$j$.

\paragraph{Action effects on beliefs.}
Defensive actions modify beliefs before propagation:
\begin{align}
\mathrm{Restore}(v)&:\;\hat{p}_v\leftarrow0.05\,\hat{p}_v,
\label{eq:restore_effect}\\
\mathrm{Remove}(v)&:\;\hat{p}_v\leftarrow0.30\,\hat{p}_v,
\label{eq:remove_effect}\\
\mathrm{Block}&:\;\alpha\leftarrow0.10\,\alpha,
\label{eq:block_effect}\\
\mathrm{Allow}&:\;\alpha\leftarrow1.05\,\alpha,
\label{eq:allow_effect}\\
\mathrm{Decoy}(v)&:\;\hat{p}_v\leftarrow0.70\,\hat{p}_v,\;
\nonumber\\
&\quad \hat{p}_w\leftarrow0.85\,\hat{p}_w\;\forall w\in\mathcal{N}(v),
\label{eq:decoy_effect}
\end{align}

where $\alpha{=}0.3$ is the default edge influence.

\paragraph{Independent cascade propagation.}
For $L{=}2$ steps:
\begin{equation}
\hat{p}_v^{(l+1)}=\max\!\left(\hat{p}_v^{(l)},\;
1-\prod_{u\in\mathcal{N}^{\mathrm{in}}(v)}(1-\alpha\,\hat{p}_u^{(l)})\right).
\label{eq:cascade}
\end{equation}
The $\max$ enforces monotonicity: compromise probability does not decrease
through propagation alone.
The superscript $(l)$ indexes the propagation depth, and
$\mathcal{N}^{\mathrm{in}}(v)$ is the set of nodes with directed influence
into $v$ under the constructed host--subnet graph.
The reported shield uses $L=2$ as a local neighbourhood screen: one hop
captures immediate subnet-to-host effects and the second hop captures the
next propagation opportunity without turning the shield into a rollout search.

\paragraph{Predicted security risk.}
With asset weights $\omega_v\in\{2,1,0.5\}$ for server/user/subnet nodes:
\begin{equation}
\hat{S}(G_t,a)
=\frac{\sum_{v\in V}\omega_v\,\hat{p}_v^{(L)}(t\mid a)}{\sum_{v}\omega_v}.
\label{eq:grp_risk}
\end{equation}
The resulting $\hat S(G_t,a)$ is a normalized post-action graph-risk score:
larger values indicate higher predicted residual compromise risk after
applying candidate action $a$ and propagating for $L$ steps.

\paragraph{Q-safe score and action selection.}
\begin{align}
Q_{\mathrm{safe}}(G_t,a)
&=-\hat{S}(G_t,a)-\sum_{k}\lambda_k\hat{C}_k(a)-\beta\,\hat{U}(G_t,a),
\label{eq:q_safe}\\
a^*&=\arg\max_{a\in\mathcal{A}_{\mathrm{safe}}(t)}Q_{\mathrm{safe}}(G_t,a),
\label{eq:grp_selection}
\end{align}
where $\hat{C}_k(a)$ uses the same cost proxies as~\eqref{eq:c_down}--\eqref{eq:c_fp},
and $\beta{=}0.1$ weights the budget-risk penalty $\hat{U}$.
For deterministic G-CRP, $\hat{U}(G_t,a)=1$ if the immediate proxy cost would
exceed any remaining budget; otherwise it is the mean normalised proxy cost
across the three budgets.  For learned G-CRP, the same symbol denotes the
predicted violation probability.
In the reported configuration $\beta=0.1$, matching the G-CRP shield code and
the ACD\textsuperscript{3} configuration.
The cascade coefficients and asset weights are hand-specified operational
priors rather than calibrated causal estimates; they define the deterministic
screen used in this study and should be recalibrated before transfer to a
different cyber range or enterprise topology.
The article reports deterministic G-CRP as the deployed action-screening rule;
the learned G-CRP fit is treated as a supervised model-fitting result, not as a
separate policy-performance claim.

\paragraph{Learned G-CRP fitting objective.}
When the learned G-CRP component is trained from collected trajectories, the
model predicts next-step process-alert probabilities, next-step network-alert
probabilities, immediate cost proxies, and violation probability.  Its
supervised loss is
\begin{equation}
\begin{aligned}
\mathcal{L}_{\mathrm{G\mbox{-}CRP}}
=&\;
\mathrm{BCE}\!\left(
\hat{\mathbf{p}}^{\mathrm{proc}},
\mathbf{p}^{\mathrm{proc}}
\right)
\;+\;
\mathrm{BCE}\!\left(
\hat{\mathbf{p}}^{\mathrm{net}},
\mathbf{p}^{\mathrm{net}}
\right)\\
&+0.1\,\lVert\hat{\mathbf{c}}-\mathbf{c}\rVert_2^2
\;+\;\mathrm{BCE}(\hat v,v),
\end{aligned}
\label{eq:learned_grp_loss}
\end{equation}
where BCE denotes binary cross-entropy, hatted quantities are model
predictions, unhatted quantities are supervised targets from the next logged
transition, $\mathbf{c}$ is the immediate cost-proxy vector, and $v$ is the
binary violation label.  This loss matches the replication-package trainer and
is reported only as a
component fit; the policy benchmark uses the deterministic G-CRP screen unless
explicitly stated otherwise.

\subsection{ACD\textsuperscript{3}-TCGS: Temporal Contract Graph Shielding}
\label{sec:tcgs}

The deterministic G-CRP screen estimates immediate graph risk from the current
observation.  Temporal Contract Graph Shielding (TCGS) adds a learned recurrent
contract-risk model that asks a different question: given the recent
trajectory history and a candidate action, how likely is the episode to cross
an operational budget within the next decision horizon?
TCGS is implemented as an opt-in diagnostic extension.  It reads existing
safety-labelled trajectories, trains a small recurrent risk model, and writes
separate evaluation traces without modifying completed checkpoints or training
outputs.

For each step, the feature vector combines the decoded binary observation,
immediate operational costs, cumulative cost normalized by the budget,
remaining budget fraction, mission phase, action-type frequencies, and scaled
reward:
\begin{equation}
\begin{aligned}
\mathbf{x}_t
=
\bigl[
&o_t \,\|\, \mathbf{c}_t
\,\|\, \mathbf{C}_t/\mathbf{B}
\,\|\, [\mathbf{B}-\mathbf{C}_t]_+/\mathbf{B}\\
&\,\|\, \mathrm{onehot}(\phi_t)
\,\|\, \mathbf{m}(a_t)
\,\|\, \bar r_t/100
\bigr]
\in\mathbb{R}^{233}.
\end{aligned}
\label{eq:tcgs_feature}
\end{equation}
Here $\mathbf{C}_t$ is the cumulative episode cost before the candidate action,
$\mathbf{B}$ is the three-budget vector, and $\mathbf{m}(a_t)$ is the
normalized action-type count vector used by the implementation.
For candidate action $a$, TCGS forms the length-$L$ history
\begin{equation}
H_t(a)=\bigl(\mathbf{x}_{t-L+1},\ldots,\mathbf{x}_{t-1},\mathbf{x}_t(a)\bigr),
\qquad L=8.
\label{eq:tcgs_history}
\end{equation}
The recurrent model is a Gated Recurrent Unit encoder with hidden dimension
96 and two heads:
\begin{equation}
\begin{aligned}
\mathbf{h}_t&=\mathrm{GRU}_\phi(H_t(a)),\\
\boldsymbol{\ell}_t&=W_v\mathbf{h}_t,\\
\hat{\mathbf{d}}_t&=\mathrm{ReLU}(W_d\mathbf{h}_t),
\end{aligned}
\label{eq:tcgs_model}
\end{equation}
Here $\boldsymbol{\ell}_t$ is the three-dimensional vector of violation logits
and $\hat{\mathbf{d}}_t$ is the non-negative predicted normalized future cost
increment.  Applying a logistic sigmoid to logit $\ell_t^k$ estimates the
probability that budget $k$ will be violated within horizon $H=100$:
\begin{equation}
p_\phi^k(H_t,a)
=\Pr_\phi\!\left[
\max_{\tau\in[t,t+H]}\sum_{s\leq \tau}c_s^k>B_k
\;\middle|\;H_t(a)
\right].
\label{eq:tcgs_prob}
\end{equation}
In Equation~\eqref{eq:tcgs_prob}, $s$ and $\tau$ are step indices inside the
future prediction window and $B_k$ is the corresponding operational budget.
The second head $\hat{\mathbf{d}}_t$ estimates the normalized future cost
increment.  The supervised training objective is
\begin{equation}
\begin{aligned}
\mathcal{L}_{\mathrm{TCGS}}
=&\;
\sum_k \mathrm{BCE}\!\left(\ell_t^k,y_t^k\right)\\
&+0.1\,\lVert \hat{\mathbf{d}}_t-\mathbf{d}_t\rVert_2^2,
\end{aligned}
\label{eq:tcgs_loss}
\end{equation}
where $y_t^k$ indicates whether budget $k$ is crossed within the prediction
horizon and $\mathbf{d}_t$ is the future normalized cost increment.

At evaluation time, a frozen ACD\textsuperscript{3}-GAT policy proposes
$a_{\mathrm{prop}}$.  TCGS accepts the proposal only if the predicted downtime
risk is below the deployability threshold and the immediate proxy cost does
not exhaust a budget:
\begin{equation}
\begin{aligned}
\chi_t(a) &=
\mathbf{1}\!\left[
p_\phi^{\mathrm{down}}(H_t,a)\leq\epsilon
\right]\\
&\quad{}\cdot
\mathbf{1}\!\left[
C_t^k+\hat c_k(a)\leq B_k,\;\forall k
\right].
\end{aligned}
\label{eq:tcgs_accept}
\end{equation}
Here $\chi_t(a)$ is an accept/reject indicator for candidate action $a$,
$C_t^k$ is the cumulative cost already incurred in the current episode, and
$\hat c_k(a)$ is the immediate cost proxy for that candidate.
\begin{equation}
a_{\mathrm{exec}}=
\begin{cases}
a_{\mathrm{prop}},&\chi_t(a_{\mathrm{prop}})=1,\\
\arg\min_{a\in\mathcal{A}_{\mathrm{valid}}}
\Psi_t(a),&\chi_t(a_{\mathrm{prop}})=0.
\end{cases}
\label{eq:tcgs_screen}
\end{equation}
with $\epsilon=0.05$ in the reported diagnostic.
If the proposal is rejected, the shield searches the valid action set and
chooses the candidate with the smallest lexicographic risk tuple $\Psi_t(a)$.
The implemented ranking
\begin{equation}
\begin{aligned}
\Psi_t(a)
&=\bigl(\psi_t^B(a),\psi_t^P(a),
p_t^{\mathrm{down}}(a),\\
&\qquad \bar p_t(a),\bar c_t(a)\bigr),\\
\psi_t^B(a)
&=\mathbf{1}[\exists k:\; C_t^k+\hat c_k(a)>B_k],\\
\psi_t^P(a)
&=\mathbf{1}[p_\phi^{\mathrm{down}}(H_t,a)>\epsilon],\\
p_t^{\mathrm{down}}(a)
&=p_\phi^{\mathrm{down}}(H_t,a),\\
\bar p_t(a)
&=\tfrac{1}{3}\sum_k p_\phi^k(H_t,a),\\
\bar c_t(a)
&=\sum_k\hat c_k(a)
\end{aligned}
\label{eq:tcgs_rank}
\end{equation}
orders candidates first by hard budget feasibility $\psi_t^B$, then by
downtime-risk threshold feasibility $\psi_t^P$, followed by predicted downtime
risk, mean predicted violation risk, and immediate proxy cost.  It evaluates
one representative valid candidate per action type for speed.  Thus TCGS is
not an end-to-end retrained policy in this
paper; it is a frozen-policy temporal shield that tests whether learned
contract-risk prediction can improve action screening online.

\subsection{Override Readiness}
\label{sec:override}

Three conditions can trigger the \textsc{Sleep} fallback:
\begin{align}
\max_a\pi(a\mid\mathrm{ctx}_t)&<\tau_{\mathrm{conf}},
\label{eq:ov_confidence}\\
\exists k:\;b_k(t)=0\;\wedge\;\hat{c}_k(a)>0,
\label{eq:ov_budget}\\
\mathrm{Std}_{a\in\mathcal{A}_{\mathrm{probe}}}\!
\left[\hat{S}(G_t,a)\right]&>\tau_{\mathrm{ood}},
\label{eq:ov_uncertainty}
\end{align}
with $\tau_{\mathrm{conf}}\in\{0.0,0.15\}$ depending on the ablation and
$\tau_{\mathrm{ood}}{=}0.7$.
Here $\tau_{\mathrm{conf}}$ is the minimum policy-confidence threshold,
$\mathcal{A}_{\mathrm{probe}}$ is the finite set of candidate actions probed
by the G-CRP uncertainty check, and $\tau_{\mathrm{ood}}$ is the threshold on
the standard deviation of predicted graph risk across those probes.
The override burden $J_{\mathrm{override}}=\mathbb{E}[\sum_t\mathbf{1}[\text{override}]]$
is penalised via $\mu\hat{A}_t^{\mathrm{override}}$ in~\eqref{eq:acd3_advantage}.

\subsection{Training Algorithm}

Algorithm~\ref{alg:acd3} writes the training horizon as
$E_{\mathrm{train}}$ because the experiments use different horizons
(30, 100, and 200 episodes in the main benchmark, with 300-episode
replications for the longer-horizon robustness check).

\begin{algorithm*}[!t]
\small
\caption{ACD$^3$-GAT Training}
\label{alg:acd3}
\begin{algorithmic}[1]
\State Initialise $\theta,\phi_r,\{\phi_{c_k}\},\phi_{\mathrm{CVaR}},
       \phi_{\mathrm{expl}},\boldsymbol\lambda=\mathbf{0}$
\For{episode $e=1,\ldots,E_{\mathrm{train}}$}
  \State Reset; $\mathbf{z}_0^i\leftarrow\mathbf{0}_{32}$;
         $b_k(0)\leftarrow B_k$; batch $\leftarrow\emptyset$
  \For{$t=0,\ldots,499$}
    \State $G_t^i\leftarrow$\textsc{ParseObs}$(o_t^i)$;\;
           $\mathbf{g}_t^i\leftarrow$\textsc{GAT}$(G_t^i)$;\;
           $\mathbf{z}_t^i\leftarrow$\textsc{GRU}$(\mathbf{g}_t^i,\mathbf{z}_{t-1}^i)$
    \State $u_t\leftarrow H(\mathrm{softmax}(W_u\mathbf{g}_t^i+\mathbf{d}_u))/\log9$;\;
           $\mathrm{ctx}_t^i\leftarrow[\mathbf{g}_t^i\|\mathbf{z}_t^i\|\mathbf{b}_t\|\mathbf{u}_t]$
    \State Sample $a_t^i\sim\pi_\theta(\cdot\mid\mathrm{ctx}_t^i)$
    \If{confidence trigger fires \eqref{eq:ov_confidence}}
      \State $a_t^i\leftarrow\textsc{Sleep}$
    \Else
      \State $a_t^i\leftarrow\mathrm{Shield}(a_t^i,G_t,\mathcal{A}_{\mathrm{safe}}(t))$
    \EndIf
    \State Execute joint action; observe $r_t^i$, $c_t^{i,k}$;\;
           $c_t^k\leftarrow\sum_i c_t^{i,k}$;\;
           $b_k(t{+}1)\leftarrow\max(0,b_k(t)-c_t^k)$
  \EndFor
  \State batch$\leftarrow$batch$\cup\{e\}$
  \If{$|\text{batch}|=8$}
    \State Compute $\hat{A}_t^{\mathrm{ACD}^3}$ via
           Eq.~\eqref{eq:acd3_advantage}; normalise
    \State Run PPO update (4 epochs over the concatenated episode batch)
    \State $\lambda_k\leftarrow[\lambda_k+0.01(\hat{J}_{c_k}-B_k)]_+$
    \State batch$\leftarrow\emptyset$
  \EndIf
  \If{adaptive Red-process evaluation enabled and $e\bmod I_{\mathrm{exploit}}=0$}
    \State Red mini-loop; update $\bar\delta$
  \EndIf
\EndFor
\end{algorithmic}
\end{algorithm*}



\section{Experiments}
\label{sec:experiments}

\subsection{Study Design}

The empirical study is organised around three questions that follow directly
from the safety-contract formulation:

\paragraph{Operational constraints.}
Do explicit operational constraints reduce budget violations without
eliminating security utility?

\paragraph{Graph structure.}
Does graph structure, through graph neural network (GNN) and Graph Attention
Network (GAT) encoders, improve over flat encoders in tail behaviour and mean
return?

\paragraph{Operational discipline.}
What does the gap between engineered CAGE~4 heuristics and naive reactive
rules reveal about operational discipline?

\subsection{Environment and Setup}

We use CAGE Challenge 4~\citep{cage4} with 5 CAGE ``Blue'' agents, episode length $T{=}500$,
and a finite-state machine CAGE ``Red'' process as the baseline source of
simulated malicious activity.
All agents receive uniform 210-dimensional binary observations
(padded to equal length across all agents).

\textbf{Operational budgets:} $B_{\text{down}}{=}50$, $B_{\text{fp}}{=}10$,
$B_{\text{fw}}{=}20$ per episode.
These values are fixed simulator stress-test thresholds for this study, not
universal SOC constants.  A production transfer would require selecting
$B_k$ from local governance constraints and rerunning the same safety-labelled
evaluation, ideally with a budget-sensitivity sweep.

\textbf{Reference points from prior work.}
In the default CAGE Challenge 4 (CC4) evaluation, the published benchmark table reports a
top heuristic reference of $-113\pm35$ and a top MARL reference of
$-193\pm84$~\citep{kiely2025cage4b}.
The $-101\pm36$ score corresponds to the constant-network-size reference
rather than the default CC4 column~\citep{kiely2025cage4b}.
The best Large Language Model (LLM) agent (GPT-o1-mini, role prompting)
scored $\approx{-2888}$~\citep{castro2025llm}.
We include the default CC4 values as reference lines in our figures and treat
them as prior-work references, not direct baselines, because our experiments
use a safety-labelled training/evaluation wrapper and explicit cost accounting.
The LLM score is reported only as a literature reference because the published
result does not provide action-level traces compatible with our downtime,
firewall, and false-positive safety accounting.

\textbf{Hyperparameters:} PPO clip $\epsilon{=}0.2$, $\gamma{=}0.99$,
$\lambda_{\text{GAE}}{=}0.95$, $\eta_\lambda{=}0.01$, $\alpha_{\text{CVaR}}{=}0.1$.
All learning methods use hidden dimension 64.
ACD\textsuperscript{3}-GAT uses $M{=}8$ episodes per update,
$(\beta,\eta,\mu)=(0.1,0.05,0.01)$ for CVaR, exploitability, and override
terms, and an exploitability mini-loop every 50 episodes when enabled.
ACD\textsuperscript{3}-GAT and C-MAPPO-GAT use
$c_{\mathrm{ent}}=0.005$; the other MAPPO-family baselines
(MAPPO-MLP, MAPPO-GNN, MAPPO-GAT, and CVaR-MAPPO) use
$c_{\mathrm{ent}}=0.01$.
All MAPPO-family learners use mini-batches of 64.
The comparison controls the main training horizon, PPO hyperparameters, and
hidden dimension, but it is not parameter-count matched: graph encoders and
factorised heads have different capacity from flat MLP policies.
We therefore interpret encoder and architecture rows as matched-protocol
comparisons, not as capacity-normalised dominance claims.
All baseline rows are presented as implemented policy families under the same
safety logger; prior published CAGE-4 heuristic and LLM scores are used only
to calibrate the reward scale because their public outputs do not expose the
action-level traces required for our operational-cost audit.

\textbf{Metric semantics and reporting conventions.}
Downtime cost is the undiscounted episode total
$\sum_t c_t^{\mathrm{down}}$; an episode violates the downtime contract iff
this total exceeds $B_{\mathrm{down}}=50$.
Figures may plot either this raw cost or the ratio
$\sum_t c_t^{\mathrm{down}}/B_{\mathrm{down}}$; captions state which is used.
CVaR-10\% is the empirical mean return of the worst 10\% of evaluated
episodes, not merely the 10th percentile.
The Pareto figures show the non-dominated points among observed methods, not
a continuous optimum of the non-convex MARL objective.
Confidence bands and box widths appear only when multiple seeds exist; methods
with one seed or short component-check horizons are plotted without inferential
claims and are labelled by their seed/episode counts in the table.
For rare violation rates, the episode count should be read together with the
seed count: for example, $P(\mathrm{viol}^{\mathrm{DT}})=0.003$ over 600
episodes corresponds to only a small number of observed violations and is
therefore evidence of strong empirical compliance in this benchmark, not a
formal probability guarantee for deployment.
All operational costs are audited as undiscounted totals because budgets are
governance limits, while reward learning still uses discounted returns.
All learned policies sample only from the valid CybORG action mask exposed by
the environment wrapper.  If a budget shield changes an action, the stored
trajectory records the executed action label and recomputed costs; the
replication records also store the run configuration, random seed, topology
seed when available, git commit, package freeze, and safety-labelled per-step
trajectory needed to recompute the benchmark tables.

\textbf{Evaluation protocol.}
Each comparison reports its seed count and episode horizon explicitly.
The study combines five evidence streams.
First, reward-only baselines quantify the operational failure mode:
Independent Proximal Policy Optimisation (IPPO), Multi-Agent PPO with a
multi-layer perceptron encoder (MAPPO-MLP), and MAPPO-GAT are each evaluated at a 200-episode horizon,
with IPPO and MAPPO-GAT replicated across three seeds.
Second, C-MAPPO-GAT, the constrained safety baseline introduced in this
paper, is evaluated over three
200-episode seeds to test whether Lagrangian costs and budget-aware action
screening reduce SOC safety-contract violations.
Third, ACD\textsuperscript{3}-GAT is evaluated as the integrated architecture,
combining graph attention, budget context, CVaR weighting, and override
with three 200-episode seeds.
Fourth, two additional 300-episode replications are run for MAPPO-GAT,
C-MAPPO-GAT, and ACD\textsuperscript{3}-GAT to test whether the main
safety conclusion survives a longer horizon.
Fifth, short 30-episode ACD\textsuperscript{3}+AskHuman and
ACD\textsuperscript{3}+deterministic G-CRP component runs verify that these
components operate inside the safety-contract regime and separate component
behaviour from the core replicated comparison.
This structure separates replicated comparisons from component checks and
keeps each quantitative claim tied to its corresponding evaluation horizon.

\subsection{Baselines}

We evaluate the following policy families:
\begin{itemize}
  \item \textbf{Sleep}: always-sleep lower bound
  \item \textbf{Random}: uniform random valid actions
  \item \textbf{Rule-based}: Monitor by default; Restore on proc alert; Block on persistent net alert
  \item \textbf{IA2C}: Independent Advantage Actor--Critic with MLP encoder.
    Prior method: \citep{mnih2016asynchronous}
    (Equations~\ref{eq:ia2c_policy_value}--\ref{eq:ia2c_loss})
  \item \textbf{IPPO}: Independent Proximal Policy Optimisation with MLP
    encoder.  Prior method: \citep{de2020independent}
  \item \textbf{MAPPO-MLP}: Multi-Agent PPO with a centralised critic and MLP
    encoder.  Prior method: \citep{yu2022surprising}
  \item \textbf{MAPPO-GNN}: Multi-Agent PPO with a centralised critic and
    GraphSAGE encoder.  Encoder prior: \citep{hamilton2017inductive}
  \item \textbf{MAPPO-GAT}: Multi-Agent PPO with a centralised critic and GAT
    encoder.  Encoder prior: \citep{velickovic2018graph}
    (\textbf{ours})
  \item \textbf{C-MAPPO-GAT}: Constrained MAPPO-GAT, introduced here as a
    controlled safety-contract baseline with Lagrangian cost
    learning~\citep{altman1999constrained,achiam2017constrained} and the
    configured hard budget-exhaustion fallback
    (\textbf{ours})
  \item \textbf{CVaR-MAPPO}: Conditional Value-at-Risk (CVaR)-MAPPO with episode-level CVaR reweighting,
    following the CVaR tail-risk objective~\citep{rockafellar2000optimization}
    and treated as a safety component rather than a separate
    safety-contract policy claim
  \item \textbf{ACD\textsuperscript{3}-GAT}: graph-attentive constrained response policy
    with budget context, CVaR weighting, and override mechanisms (\textbf{ours};
    three 200-episode seeds)
\end{itemize}
For MAPPO-GAT and C-MAPPO-GAT, the GAT encoder in
Equations~\ref{eq:gat_score}--\ref{eq:global_emb} instantiates
$\mathrm{Enc}_\phi^i$ in the centralised critic of
Equation~\ref{eq:mappo_critic}; the actor still executes from each agent's
local observation.  For ACD\textsuperscript{3}-GAT, the same graph embedding
$\mathbf{g}_t^i$ is concatenated with opponent, budget, and uncertainty
context in Equation~\ref{eq:ctx} before the factorised action policy and
safety shield are applied.

\subsection{Benchmark Method Instantiations}
\label{sec:benchmark_instantiations}

All learned methods use the CAGE-4 observation and action spaces defined in Section~\ref{sec:problem}, and they differ only in the policy update, encoder, critic information, and safety machinery. This subsection fixes the mapping between Table~\ref{tab:benchmark} and the implemented algorithms.

\paragraph{Non-learning reference policies.}
Sleep is the deterministic policy
$\pi(a_t^i=\textsc{Sleep}\mid o_t^i)=1$.
Random samples uniformly from the valid action mask
$\mathcal{A}_{\mathrm{valid}}^i(o_t^i)$:
\begin{equation}
\pi_{\mathrm{rand}}(a\mid o_t^i)
=|\mathcal{A}_{\mathrm{valid}}^i(o_t^i)|^{-1}
\mathbf{1}[a\in\mathcal{A}_{\mathrm{valid}}^i(o_t^i)].
\label{eq:random_policy}
\end{equation}
Here $\mathcal{A}_{\mathrm{valid}}^i(o_t^i)$ is the CybORG action mask for
agent $i$ in observation $o_t^i$, so invalid simulator actions receive zero
probability.
The rule-based policy is a deterministic alert policy: monitor by default,
restore when process alerts are present, and block traffic when persistent
network alerts are visible.  These policies use no learned parameters; their
returns and safety metrics are computed with the same logger and cost proxies
as the learned methods.

\paragraph{IA2C and IPPO.}
IA2C follows Equations~\ref{eq:ia2c_policy_value}--\ref{eq:ia2c_loss}.
IPPO uses the same independent per-agent information pattern, but replaces
the vanilla actor--critic objective with clipped PPO:
\begin{equation}
\pi_{\theta_i}(a_t^i\mid o_t^i),\quad
V_{\phi_i}(o_t^i),\quad
\mathcal{L}_{\mathrm{IPPO}}^i
=\mathcal{L}^{\mathrm{CLIP}}_i
+c_{\mathrm{vf}}\mathcal{L}^{\mathrm{VF}}_i
-c_{\mathrm{ent}}H[\pi_{\theta_i}],
\label{eq:ippo_instantiation}
\end{equation}
with GAE advantages from Equations~\ref{eq:gae_base}--\ref{eq:gae_recurrence}.
The actor parameters $\theta_i$, critic parameters $\phi_i$, local observation
$o_t^i$, and entropy/value-loss weights are the same quantities defined for
IA2C and PPO in Section~\ref{sec:foundations}.
The IPPO row uses an MLP encoder; the Factorized-IPPO row keeps the same
independent PPO update but replaces the flat actor with the type--target
factorisation in Equation~\ref{eq:factorized_policy}.

\paragraph{MAPPO encoder variants.}
MAPPO-MLP, MAPPO-GNN, and MAPPO-GAT all use the centralised critic in
Equation~\ref{eq:mappo_critic} and the clipped PPO objective in
Equation~\ref{eq:ppo_total}.  Their only architectural difference is the
per-agent encoder:
\begin{equation}
\mathrm{Enc}_\phi^i(o_t^i)=
\begin{cases}
\mathrm{MLP}(o_t^i), & \text{MAPPO-MLP},\\
\mathrm{GraphSAGE}(G_t^i), & \text{MAPPO-GNN},\\
\mathrm{GAT}(G_t^i), & \text{MAPPO-GAT}.
\end{cases}
\label{eq:mappo_encoder_instantiation}
\end{equation}
Here $\mathrm{Enc}_\phi^i$ denotes the per-agent encoder whose output is sent
to the centralised critic during training; $G_t^i$ is the graph parsed from
the same local observation $o_t^i$.
GraphSAGE and GAT are defined in
Equations~\ref{eq:graphsage_update} and
\ref{eq:gat_score}--\ref{eq:global_emb}, respectively.
Actors execute decentralised local policies, while the critic uses the
concatenated five-agent embedding during training.

\paragraph{Opponent-conditioned IPPO.}
Opp-IPPO keeps the independent PPO loss but augments each agent's policy state
with a recurrent opponent embedding:
\begin{equation}
\begin{aligned}
\mathbf{z}_t^i
&=\mathrm{GRU}(\mathrm{Enc}(o_t^i),\mathbf{z}_{t-1}^i),\\
a_t^i
&\sim\pi_{\theta_i}(\cdot\mid o_t^i,\mathbf{z}_t^i).
\end{aligned}
\label{eq:opp_ippo_instantiation}
\end{equation}
Here $\mathbf{z}_t^i$ is the recurrent state for agent $i$, initialised to
zero at the start of each episode and updated from that agent's encoded local
observation.
This is the same recurrent mechanism later reused in the ACD\textsuperscript{3}
context vector, but without safety budgets, CVaR weighting, or shielding.

\paragraph{CVaR-MAPPO.}
CVaR-MAPPO keeps the MAPPO centralised critic and PPO update, but multiplies
the policy advantage by an episode-level tail weight.  For a batch of $M$
episodes and tail fraction $\alpha=0.1$,
\begin{equation}
\begin{aligned}
k^* &= \max(1,\lfloor\alpha M\rfloor),\\
w_e &=
\begin{cases}
1/k^*, & e\in\text{worst }k^*\text{ episodes},\\
0, & \text{otherwise},
\end{cases}\\
\hat A_t^{\mathrm{CVaR\mbox{-}MAPPO}}
&=w_e\hat A_t^{\mathrm{MAPPO}} .
\end{aligned}
\label{eq:cvar_mappo_instantiation}
\end{equation}
Here $e$ indexes episodes in the PPO batch, $w_e$ is the binary tail weight
assigned to the worst-return episodes, and
$\hat A_t^{\mathrm{MAPPO}}$ is the reward-only MAPPO advantage.
This row therefore tests tail-risk reweighting alone, without Lagrange
constraints or action shielding.

\paragraph{Constrained MAPPO-GAT.}
C-MAPPO-GAT is our constrained MAPPO-GAT instantiation: it keeps the MAPPO-GAT
encoder and centralised critic, adds cost-advantage penalties for the three
SOC budgets, and applies the configured hard budget-exhaustion fallback:
\begin{equation}
\begin{aligned}
\hat A_t^{\mathrm{C\mbox{-}MAPPO}}
&=\hat A_t^{r}
-\sum_k\lambda_k\hat A_t^{c_k},\\
\lambda_k
&\leftarrow
[\lambda_k+\eta_\lambda(\hat J_{c_k}-B_k)]_+ .
\end{aligned}
\label{eq:constrained_mappo_instantiation}
\end{equation}
Here $\hat A_t^r$ is the reward advantage, $\hat A_t^{c_k}$ is the advantage
computed from cost stream $k$, $\lambda_k$ is the corresponding Lagrange
multiplier, and $\hat J_{c_k}$ is the observed batch cost estimate compared
with budget $B_k$.
This is the safety-contract baseline against which ACD\textsuperscript{3}-GAT
is interpreted.

\paragraph{ACD\textsuperscript{3} variants.}
ACD\textsuperscript{3}-GAT uses the full context vector
Equation~\ref{eq:ctx}, factorised policy Equation~\ref{eq:factorized_policy},
composite advantage Equation~\ref{eq:acd3_advantage}, and shielded execution
Equation~\ref{eq:shielded_policy}.
The AskHuman and deterministic G-CRP rows are short configuration checks of
the override and action-screening components inside the same ACD\textsuperscript{3}
scaffold; they are not presented as separately optimised policy families.

\subsection{G-CRP Shield Diagnostics}

To isolate the contribution of Graph Counterfactual Risk Propagation
(Section~\ref{sec:grp}), we define three shield variants while holding the
policy class (ACD\textsuperscript{3}-GAT) fixed:

\paragraph{Rule-based shield.}
The reactive shield blocks actions whose cost proxy is positive for an
exhausted budget; it uses no prediction and no graph reasoning.

\paragraph{Deterministic G-CRP.}
The physics-based shield uses the cascade propagation model in
Equation~\ref{eq:cascade}, selects $\arg\max Q_{\text{safe}}$
(Equation~\ref{eq:q_safe}), and requires no training.

\paragraph{Learned G-CRP.}
The supervised extension trains a GraphSAGE model on collected trajectories to
predict next-step alert state, cost, and violation probability before action
screening.

The deterministic G-CRP path is included as part of the
ACD\textsuperscript{3} action-screening design and is evaluated in a short
configuration check.
The learned G-CRP model is trained in the replication package as the supervised
risk-estimation extension of the deterministic screen.
The quantitative policy comparison reports the deterministic G-CRP path so
that each policy-performance claim is tied to the action-screening rule used
in the evaluated runs.
When that component is evaluated, the intended measures are violation
rate, mean return, override rate, violation-probability calibration, and
cost-prediction error.

\subsection{Robustness Extensions}

The robustness evaluation tests whether the safety contract survives outside
the matched training and evaluation seed setting:
\begin{itemize}
  \item \textbf{In-distribution (ID)}: 5 topology seeds used during training
  \item \textbf{Topology out-of-distribution (OOD)}: 5 unseen seeds (different host counts)
  \item \textbf{Red-process OOD protocol}: a separate evaluation with Red policies
    not used in Blue training or tuning, reserved for future quantitative
    claims beyond the present split-labelled seed-variation artifacts
  \item \textbf{Mission OOD}: altered mission phase schedule
  \item \textbf{Adaptive Red process}: stress curve -- Red PPO trained 20 rounds against frozen Blue policies
\end{itemize}

The present benchmark reports the ID setting and two robustness checks.
First, IPPO, MAPPO-GAT, and constrained MAPPO-GAT are evaluated on unseen
topology seeds for 20 episodes per seed, using two seeds per split and the
same finite-state Red process.
Second, a Red PPO league is trained for 20 updates against frozen Blue
policies, with sampled Red-process actions coupled into the CybORG transition loop.
The split-labelled Red-process artifacts collected in this replication package
use the same CAGE finite-state Red-process family with seed variation and are
therefore not interpreted as held-out Red-policy evidence.
Held-out Red-process and mission-schedule variants remain part of the
evaluation protocol but are not used as quantitative claims here.



\section{Results}
\label{sec:results}

\subsection{Evaluation Context}

All experiments use CAGE Challenge~4~\citep{cage4} with $N{=}5$ CAGE ``Blue''
agents, $T{=}500$ steps, and the finite-state machine CAGE ``Red'' process.
All agents receive a uniform 210-dimensional binary observation vector.
Operational budgets: $B_{\mathrm{down}}{=}50$, $B_{\mathrm{fp}}{=}10$,
$B_{\mathrm{fw}}{=}20$ per episode.
The main replicated comparison uses three 200-episode seeds for IPPO,
MAPPO-GAT, constrained MAPPO-GAT, and ACD\textsuperscript{3}-GAT.
MAPPO-MLP provides a one-seed flat-encoder reference at the same horizon.
For the three principal graph-based methods, two additional 300-episode
replications assess whether the central safety ordering persists when training
is extended.
Shorter architectural component runs (100 or 30 episodes) are used only to
interpret component behaviour and are labelled separately from the replicated
comparison; they are not used to claim a new best policy.

\subsection{Safety and Return}
\label{sec:main_results}

\begin{figure*}[!t]
\centering
\includegraphics[width=0.95\textwidth]{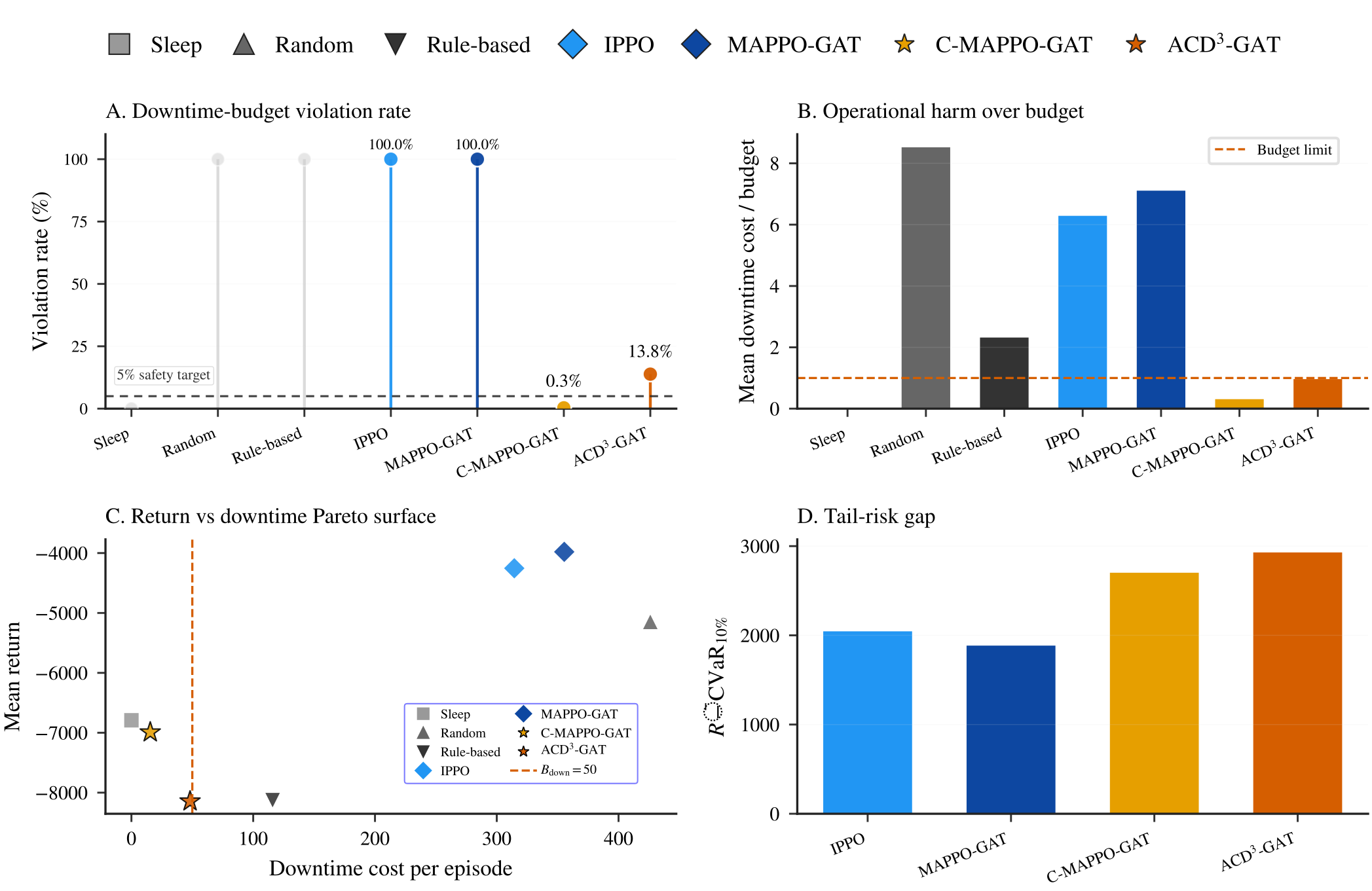}
\caption{Main empirical story in four focused panels.  The colors and markers
  identify the representative methods shown in the dashboard; the complete
  benchmark, including IA2C, MAPPO-MLP/GNN, Factorized-IPPO, Opp-IPPO, and
  CVaR-MAPPO, is reported in Table~\ref{tab:benchmark}.
  \textbf{(a)} Reward-only policies violate the downtime budget in nearly
  every episode; C-MAPPO-GAT is the most reliable safety-contract
  configuration, while ACD\textsuperscript{3}-GAT is evaluated as the
  integrated safety-contract architecture.
  \textbf{(b)} The same methods consume multiple MTTR budgets per episode;
  the dashed boundary corresponds to
  $\sum_t c_t^{\mathrm{down}}/B_{\mathrm{down}}=1$.
  \textbf{(c)} The return--downtime Pareto surface exposes the operational
  price of constraint compliance and the current ACD\textsuperscript{3}-GAT
  frontier position.
  \textbf{(d)} The gap between mean return and CVaR-10\% shows that tail
  episodes remain a first-class risk even when average reward improves.}
\label{fig:story_dashboard}
\end{figure*}

\begin{figure*}[!t]
\centering
\begin{minipage}{0.48\textwidth}
  \centering
  \includegraphics[width=\linewidth]{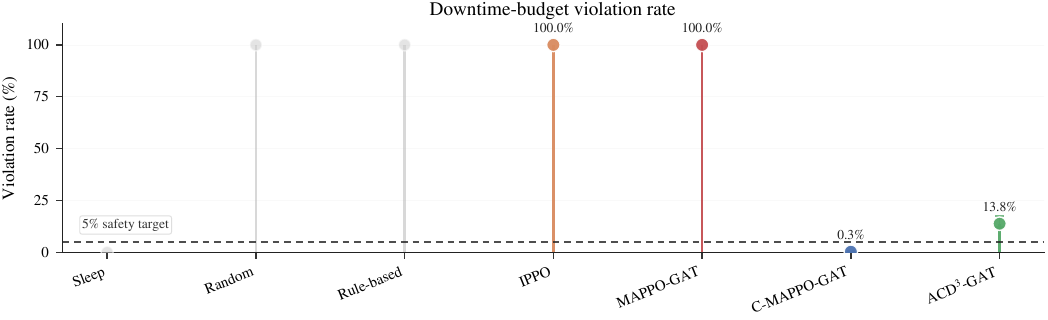}
  \vspace{-2pt}
  {\footnotesize\textbf{(a)} Downtime-budget violation.}
\end{minipage}\hfill
\begin{minipage}{0.48\textwidth}
  \centering
  \includegraphics[width=\linewidth]{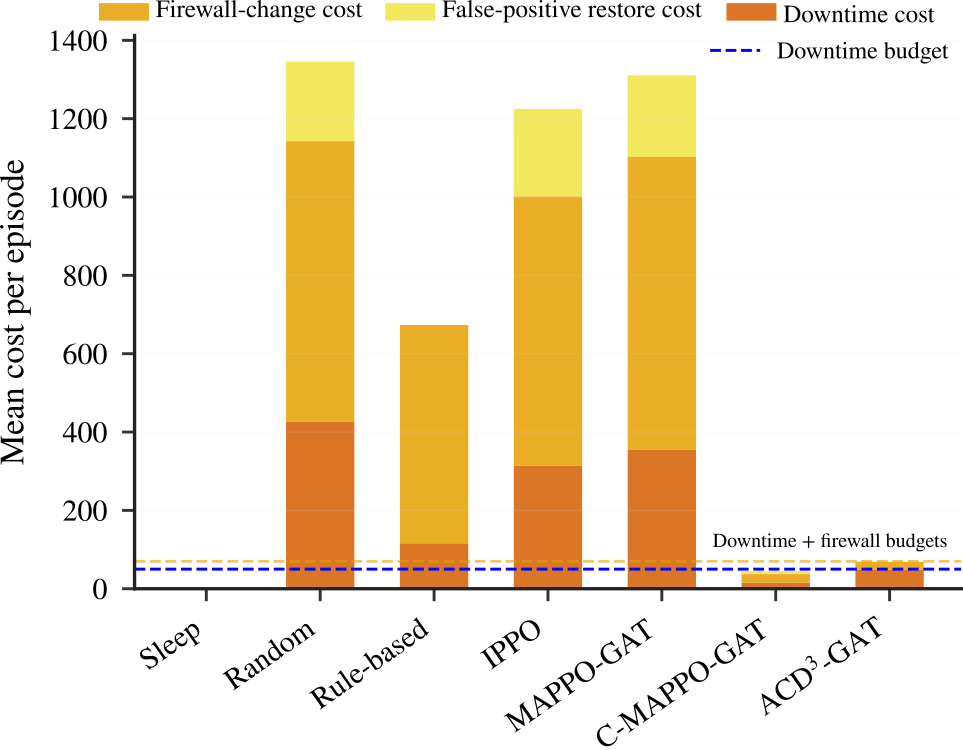}
  \vspace{-2pt}
  {\footnotesize\textbf{(b)} Operational cost decomposition.}
\end{minipage}

\vspace{6pt}

\begin{minipage}{0.48\textwidth}
  \centering
  \includegraphics[width=\linewidth]{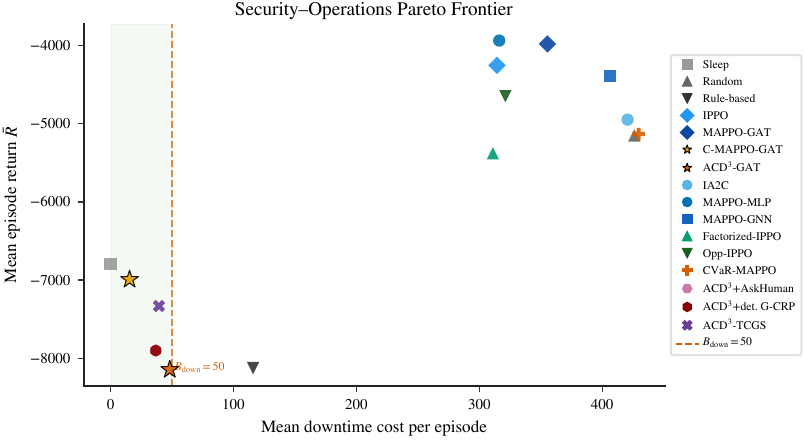}
  \vspace{-2pt}
  {\footnotesize\textbf{(c)} Return--downtime frontier.}
\end{minipage}\hfill
\begin{minipage}{0.48\textwidth}
  \centering
  \includegraphics[width=\linewidth]{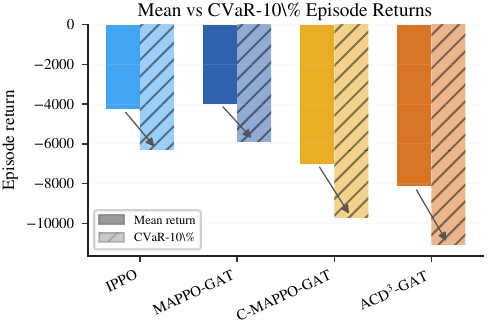}
  \vspace{-2pt}
  {\footnotesize\textbf{(d)} Mean return and CVaR-10\%.}
\end{minipage}
\caption{Safety-contract benchmark decomposed into violation, operational
cost, return--cost frontier, and tail-risk views.
\textbf{(a)} A violation is the episode-level event
$\sum_t c_t^{\mathrm{down}}>B_{\mathrm{down}}$; the dashed line marks a
5\% safety-contract target.
\textbf{(b)} Downtime (Restore), firewall churn (Block/Allow), and
false-positive Restore costs are undiscounted episode totals.
\textbf{(c)} The vertical boundary marks the downtime budget
$B_{\mathrm{down}}=50$; points left of the boundary satisfy the mean MTTR
budget.
\textbf{(d)} CVaR-10\% is the mean return of the worst 10\% of episodes,
showing that reward improvements do not automatically remove worst-tail
episodes.}
\label{fig:benchmark_decomposition}
\end{figure*}

\begin{figure*}[!t]
\centering
\includegraphics[width=0.95\textwidth]{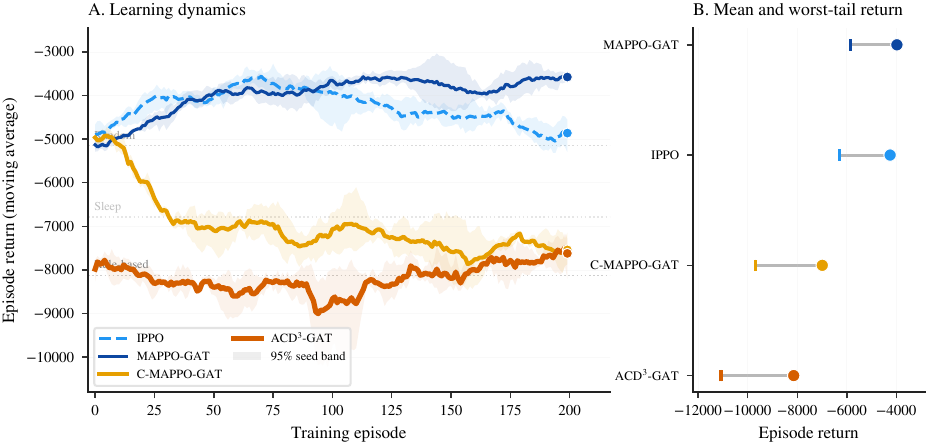}
\caption{Learning dynamics and final return-risk context for the principal learning
  methods.
  \textbf{(a)} Moving-average episode returns over the common 200-episode
  benchmark window, with shaded seed bands for replicated methods.
  Non-learning baselines are shown only as muted scale references.
  \textbf{(b)} Final mean return and CVaR-10\% for the same methods.
  This pairing should be read with the violation figures because reward alone
  is not the deployment criterion.}
\label{fig:learning_curves}
\end{figure*}

\begin{figure*}[!t]
\centering
\includegraphics[width=0.95\textwidth]{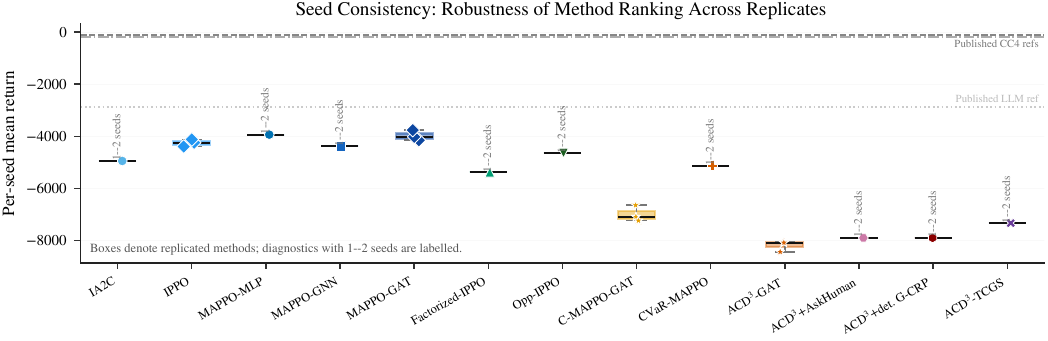}
\caption{Seed-consistency audit for learning methods.
  Each point is one seed's mean episode return and boxes summarise the
  across-seed distribution.  Single-seed and short component-check methods are
  explicitly marked, separating replicated comparisons from exploratory
  component checks.}
\label{fig:seed_consistency}
\end{figure*}

Table~\ref{tab:benchmark} summarises the safety-contract benchmark.
The table reports the complete direct comparison under our safety-labelled
evaluation protocol: it includes the non-learning baselines, IA2C, IPPO, MAPPO with
MLP/GraphSAGE/GAT encoders, factorised and opponent-conditioned IPPO variants,
CVaR-MAPPO, C-MAPPO-GAT, and ACD\textsuperscript{3}-GAT.
The figures provide the same evidence at different
levels of resolution.
Figure~\ref{fig:story_dashboard} summarizes the benchmark through four linked views: budget violation, budget overrun, return--downtime tradeoff, and tail-risk gap.
Figure~\ref{fig:benchmark_decomposition} then unpacks that dashboard into
four benchmark views: panel~(a) asks whether the episode exceeds
$B_{\mathrm{down}}$; panel~(b) identifies which action-derived costs are
responsible for the separation; panel~(c) places the same methods on a
return--downtime plane; and panel~(d) checks whether mean return is hiding
poor worst-tail outcomes.
Figure~\ref{fig:learning_curves} connects the final summaries to training
dynamics rather than treating the table as a single endpoint, and
Figure~\ref{fig:seed_consistency} makes the replication status visible by
separating three-seed comparisons from one-seed or short component checks.
Finally, Figure~\ref{fig:exploitability} reports the coupled adaptive
Red-process stress test, while Figure~\ref{fig:grp_physics} visualises the
deterministic graph counterfactual model used by the shield.
Published CAGE-4 and LLM numbers appear only as prior-work reference lines or
table notes; they are not treated as direct baselines because they lack the
same safety-labelled action traces and budget accounting.

\begin{table*}[!t]
\centering
\caption{Benchmark summary for the safety-contract evaluation. Episode and seed columns make the replication depth explicit. ``Ep.'' is the total number of evaluated episodes included for each row. $\bar{R}$ is mean episode return; $\mathrm{CVaR}_{10\%}$ is the worst-tail mean return averaged over per-run metrics; $P(\mathrm{viol}^{\mathrm{DT}})$ is downtime-budget violation rate; $\bar{c}^{\mathrm{DT}}$ is mean downtime cost; Catast. is the catastrophic episode rate under the alert-threshold proxy. $\dagger$ marks the most conservative safety-contract compliance row in this benchmark; $\star$ marks the integrated ACD\textsuperscript{3}-GAT method architecture on the broader safety-contract frontier. Citations in the method column identify prior algorithmic families; C-MAPPO-GAT and ACD\textsuperscript{3}-GAT are introduced here as safety-contract instantiations built from those components.}
\label{tab:benchmark}
\begingroup
\footnotesize
\setlength{\tabcolsep}{1.8pt}
\renewcommand{\arraystretch}{1.12}
\begin{tabularx}{\textwidth}{@{}>{\raggedright\arraybackslash}p{0.105\textwidth}>{\raggedright\arraybackslash}X>{\centering\arraybackslash}p{0.043\textwidth}>{\centering\arraybackslash}p{0.045\textwidth}>{\raggedleft\arraybackslash}p{0.066\textwidth}>{\raggedleft\arraybackslash}p{0.066\textwidth}>{\centering\arraybackslash}p{0.053\textwidth}>{\raggedleft\arraybackslash}p{0.058\textwidth}>{\centering\arraybackslash}p{0.047\textwidth}@{}}
\toprule
Group & Method & Ep. & S & $\bar{R}$ & CVaR & $P_{\mathrm{DT}}$ & $\bar{c}_{\mathrm{DT}}$ & Cat. \\
\midrule
\textit{Nonlearn.} & Sleep & 140 & 3 & $-6{,}792$ & $-8{,}984$ & 0.000 & 0.0 & 0.000 \\
 & Random & 140 & 3 & $-5{,}149$ & $-6{,}737$ & 1.000 & 426.1 & 0.107 \\
 & Rule-based~\citep{kiely2025cage4} & 140 & 3 & $-8{,}124$ & $-10{,}456$ & 1.000 & 115.9 & 0.000 \\
\addlinespace
\textit{Actor-critic} & IA2C~\citep{mnih2016asynchronous} & 100 & 1 & $-4{,}948$ & $-6{,}691$ & 1.000 & 420.6 & 0.070 \\
 & IPPO~\citep{schulman2017proximal} & 600 & 3 & $-4{,}254$ & $-6{,}299$ & 1.000 & 314.4 & 0.012 \\
\addlinespace
\textit{MAPPO enc.} & MAPPO-MLP~\citep{yu2022surprising} & 200 & 1 & $-3{,}937$ & $-6{,}226$ & 1.000 & 316.2 & 0.015 \\
 & MAPPO-GNN~\citep{yu2022surprising,hamilton2017inductive} & 100 & 1 & $-4{,}387$ & $-6{,}108$ & 1.000 & 406.2 & 0.010 \\
 & MAPPO-GAT~\citep{yu2022surprising,velickovic2018graph} & 600 & 3 & $-3{,}979$ & $-5{,}864$ & 1.000 & 355.4 & 0.035 \\
\addlinespace
\textit{Arch.} & Fact.-IPPO~\citep{schulman2017proximal} & 100 & 1 & $-5{,}378$ & $-7{,}428$ & 1.000 & 311.1 & 0.120 \\
 & Opp-IPPO~\citep{schulman2017proximal,cho2014learning} & 100 & 1 & $-4{,}648$ & $-7{,}015$ & 1.000 & 321.2 & 0.000 \\
\addlinespace
\textit{Safety} & CVaR-MAPPO~\citep{yu2022surprising,rockafellar2000optimization} & 100 & 1 & $-5{,}131$ & $-6{,}509$ & 1.000 & 429.6 & 0.090 \\
 & \textbf{C-MAPPO-GAT}~\citep{yu2022surprising,velickovic2018graph,altman1999constrained} $\dagger$ & 600 & 3 & $-6{,}992$ & $-9{,}694$ & 0.003 & 15.5 & 0.000 \\
 & ACD\textsuperscript{3}-GAT~\citep{yu2022surprising,velickovic2018graph,altman1999constrained,rockafellar2000optimization} $\star$ & 600 & 3 & $-8{,}144$ & $-11{,}074$ & 0.138 & 48.2 & 0.003 \\
\addlinespace
\textit{ACD\textsuperscript{3} comp.} & ACD\textsuperscript{3}+AskHuman & 30 & 1 & $-7{,}901$ & $-10{,}460$ & 0.100 & 36.8 & 0.000 \\
 & ACD\textsuperscript{3}+det. G-CRP & 30 & 1 & $-7{,}901$ & $-10{,}460$ & 0.100 & 36.8 & 0.000 \\
\bottomrule
\end{tabularx}
\endgroup
\end{table*}


\noindent\textit{Component-check rows.}
The final two ACD\textsuperscript{3} rows in Table~\ref{tab:benchmark}
are implementation checks of the governance layer rather than separate
policy claims.
Both reuse the same trained ACD\textsuperscript{3}-GAT checkpoint for a
30-episode, single-seed check while emphasizing a different screening path:
the AskHuman escalation gate or the deterministic G-CRP shield.
Under the default confidence setting ($\tau_{\mathrm{conf}}=0.0$), the
AskHuman gate does not separate the aggregate metrics in this short window,
so the two rows coincide.
They are included to show that these screening hooks preserve the learned
reward and compliance profile under the short component check; attributing
performance differences between the hooks would require a longer targeted
evaluation.


Read together, Table~\ref{tab:benchmark} and
Figure~\ref{fig:story_dashboard} establish the central empirical pattern.
In the dashboard's compliance panel, the unconstrained learners cluster at
full downtime-budget violation, meaning that the policies cross
$B_{\mathrm{down}}$ in every evaluated episode even when their reward is
competitive.
The adjacent over-budget panel explains the scale of the failure: the same
methods consume several multiples of the allowed MTTR budget, whereas
C-MAPPO-GAT moves below the budget and ACD\textsuperscript{3}-GAT moves close
to it on average.
The Pareto panel then shows the price of that movement.
Reward-only MAPPO-GAT and random exploration sit high in return but far to
the right of the downtime boundary; C-MAPPO-GAT moves into the feasible
operational region at a return cost; and ACD\textsuperscript{3}-GAT occupies
an intermediate frontier point that reflects the integrated architecture
rather than the most conservative compliance setting.
The final dashboard panel makes the same argument for tail behaviour, which is that the gap between mean return and CVaR-10\% remains large enough that average reward cannot serve as the sole deployment criterion.


\paragraph{Constraint compliance.}
C-MAPPO-GAT reaches $P(\mathrm{viol}^{\mathrm{DT}}) = 0.003$ across
600 episodes from three seeds.
This corresponds to two downtime-budget violations in the replicated core sample, providing empirical compliance evidence under the reported benchmark but not a certified chance constraint.
ACD\textsuperscript{3}-GAT reaches
$P(\mathrm{viol}^{\mathrm{DT}}) = 0.138$ with mean downtime cost $48.2$
across three 200-episode seeds: below the episode budget on average, but with
less reliable episode-level compliance than C-MAPPO-GAT.
Every unconstrained learning method in Table~\ref{tab:benchmark} achieves
$P(\mathrm{viol}^{\mathrm{DT}}) = 1.000$.
The 200-episode IPPO/MAPPO rows still consume 314--355 downtime-cost
units per episode, while constrained MAPPO-GAT consumes \textbf{15.5}, a
\textbf{95--96\% reduction} against a budget of 50.
ACD\textsuperscript{3}-GAT consumes \textbf{48.2}, an \textbf{84--86\%}
reduction in mean downtime cost relative to the same unconstrained policies,
with a 13.8\% violation rate.
This places the integrated policy on the safety-contract frontier rather than
at the most conservative compliance point: it demonstrates the full
ACD\textsuperscript{3} architecture, while C-MAPPO-GAT---introduced here as
the constrained MAPPO/GAT/Lagrangian instantiation of the same safety-contract
idea---represents the strongest compliance configuration in the present
benchmark.

The cost savings do not come for free: constrained MAPPO-GAT and
ACD\textsuperscript{3}-GAT both have lower return than the unconstrained
MAPPO variants.
The resulting Pareto tradeoff is explicit: operational safety costs security
reward, but a policy with $P(\mathrm{viol}^{\mathrm{DT}})=1.000$ cannot be
considered deployable under the stated SOC contract.
Figure~\ref{fig:benchmark_decomposition} decomposes this distinction across violation probability, operational cost, return--downtime tradeoff, and tail-risk gap.
Panel~(a) turns downtime into a governance event--crossing the episode
budget--rather than a continuous training metric.
Panel~(c) then shows why the safety result is not simply a lower-score variant
of the same policy: the constrained methods move the system into a different
operating regime, where lower reward is exchanged for remaining inside or near
the MTTR budget.
Panel~(b) explains that downtime Restore cost is the dominant operational harm
separating the methods, while firewall-change and false-positive costs remain
important secondary contract dimensions.
Panel~(d) keeps the interpretation honest: even when a method improves mean
return, its worst episodes can remain much worse than its average behaviour.

\paragraph{Graph structure and tail behaviour.}
In the benchmark, MAPPO-MLP at 200 episodes
($-3{,}937$) remains ahead of MAPPO-GAT across 600 episodes
($-3{,}979$), while MAPPO-GAT has the strongest CVaR-10\% score among the
unconstrained MAPPO encoder variants.
The evidence therefore supports a narrower conclusion:
graph structure improves tail behaviour, but the raw-return
advantage of graph attention has not emerged in the three-seed
MAPPO-GAT comparison.
Figure~\ref{fig:learning_curves} is useful here because it shows that the
reward-only policies learn quickly toward higher return while remaining
operationally unsafe, whereas the constrained curves occupy a lower-return
region because the policy update is also paying for downtime, firewall, and
false-positive costs.
Figure~\ref{fig:seed_consistency} complements this by showing the spread of
seed-level returns and marking which rows are replicated enough to support
the core comparison.
The visual message is therefore not that graph attention dominates every
metric, but that graph-aware and constrained variants expose different axes of
performance: representation affects tail return, while operational
constraints affect deployability.


\paragraph{Longer-horizon replication.}
Table~\ref{tab:long_horizon} reports the two additional 300-episode
replications for MAPPO-GAT, constrained MAPPO-GAT, and
ACD\textsuperscript{3}-GAT.
The longer horizon preserves the main ordering.
Reward-only MAPPO-GAT remains operationally non-deployable:
$P(\mathrm{viol}^{\mathrm{DT}})=1.000$ and mean downtime cost 357.1,
despite strong mean return ($\bar{R}=-4{,}033$).
Constrained MAPPO-GAT remains the most reliable safety-contract policy, with
$P(\mathrm{viol}^{\mathrm{DT}})=0.007$, mean downtime cost 10.1, and zero
catastrophic episodes across the two 300-episode replications.
ACD\textsuperscript{3}-GAT remains close to the downtime budget on average
($\bar{c}^{\mathrm{DT}}=48.8$), with a 14.3\% episode-violation rate.
The extension strengthens the central safety-contract result and preserves
the same frontier interpretation:
the integrated method reduces operational harm, while the constrained
baseline remains the most conservative compliance configuration in the
reported evaluations.

\begin{table*}[!t]
\centering
\caption{Longer-horizon replication check for the three principal graph-based
learning methods.  The table reports two additional 300-episode replications
per method, evaluated separately from the balanced three-seed benchmark in
Table~\ref{tab:benchmark}.}
\label{tab:long_horizon}
\begingroup
\scriptsize
\setlength{\tabcolsep}{2pt}
\renewcommand{\arraystretch}{1.12}
\begin{tabularx}{\textwidth}{@{}>{\raggedright\arraybackslash}X>{\centering\arraybackslash}p{0.05\textwidth}>{\centering\arraybackslash}p{0.05\textwidth}>{\raggedleft\arraybackslash}p{0.075\textwidth}>{\raggedleft\arraybackslash}p{0.075\textwidth}>{\centering\arraybackslash}p{0.06\textwidth}>{\raggedleft\arraybackslash}p{0.065\textwidth}>{\centering\arraybackslash}p{0.05\textwidth}@{}}
\toprule
Method & Ep. & S & $\bar{R}$ & CVaR & $P_{\mathrm{DT}}$ & $\bar{c}_{\mathrm{DT}}$ & Cat. \\
\midrule
MAPPO-GAT & 600 & 2 & $-4{,}033$ & $-5{,}848$ & 1.000 & 357.1 & 0.063 \\
C-MAPPO-GAT & 600 & 2 & $-6{,}962$ & $-9{,}594$ & 0.007 & 10.1 & 0.000 \\
ACD\textsuperscript{3}-GAT & 600 & 2 & $-8{,}132$ & $-11{,}082$ & 0.143 & 48.8 & 0.002 \\
\bottomrule
\end{tabularx}
\endgroup
\end{table*}


\subsection{Operational Discipline and Heuristic Baselines}
\label{sec:heuristic_failure}

A central finding from the CAGE4 competition~\citep{kiely2025cage4} is that
\emph{engineered heuristic agents outperformed the submitted MARL agents}.
Our local rule-based baseline is not a reproduction of the winning heuristic.
It is a deliberately simple reactive policy that isolates the role of
operational discipline.

Our rule-based agent achieves $\bar{R} \approx -8{,}124$ --
\emph{worse than doing nothing} (Sleep: $\approx -6{,}792$) and far below random
($\approx -5{,}149$).
The mechanism is clear from the cost breakdown:
$\bar{c}^{\mathrm{DT}} \approx 115$ (rule-based) vs $0$ (sleep) and $425$ (random).
The rule agent triggers \textsc{Restore} on every malicious-process alert,
eventually exhausting $B_{\mathrm{down}} = 50$ because the cost proxy counts
one unit per Restore action.
Each subsequent Restore action is a budget violation, and the accumulated cost
overwhelms the security benefit.

This motivates constrained learning:
naive rules fail because they lack the observation engineering, valid-action
filtering, mission-phase traffic discipline, and selective response logic that
made the top CAGE4 heuristics effective.
Unconstrained RL fails for a complementary reason: it discovers high-impact
actions such as \textsc{Restore} but does not internalise the operational budget.
The target for any deployable autonomous response policy is therefore not
just high security reward, but \emph{safety-contract compliance}.

\subsection{Architectural and Component Analysis}
\label{sec:ablations}

\paragraph{Encoder architectures.}
The MAPPO family covers three encoder architectures.
At the reported horizons:
\begin{itemize}
  \item \textbf{MLP}: $\bar{R} = -3{,}937$, $\mathrm{CVaR}_{10\%} = -6{,}226$
        at 200 episodes. Best raw return; no structural inductive bias.
  \item \textbf{GNN} (GraphSAGE): $\bar{R} = -4{,}387$,
        $\mathrm{CVaR}_{10\%} = -6{,}108$ at 100 episodes.
        Graph aggregation reduces tail episodes.
  \item \textbf{GAT}: $\bar{R} = -3{,}979$, $\mathrm{CVaR}_{10\%} = -5{,}864$
        across 600 episodes from three seeds. Best reported MAPPO
        encoder CVaR-10\%;
        The attention mechanism adds interpretability at modest extra cost.
\end{itemize}
The graph encoders improve the tail metric relative to the MLP run
($-6{,}108$ for GNN and $-5{,}864$ for GAT vs $-6{,}226$ for MLP), while
the MLP still leads in mean return.
In this benchmark, graph structure contributes primarily to tail-risk behaviour
and interpretability rather than to raw-return dominance.

\paragraph{Factorized action head.}
Factorized-IPPO ($-5{,}378$ at 100 episodes) underperforms replicated IPPO
($-4{,}254$ across 600 episodes).
The factorized head (type $\times$ target) has more parameters and slower
apparent convergence in this short 100-episode component run, especially with
the GAT encoder.
However, its catastrophic episode rate (0.120 vs 0.012 for IPPO) is higher,
so the factorized variant remains an instability to revisit in longer ablations.

\paragraph{Opponent embedding.}
Opp-IPPO ($-4{,}648$ at 100 episodes) trails replicated IPPO
($-4{,}254$ across 600 episodes) in return but achieves catastrophic
rate 0.000 vs 0.012 for IPPO.
This is consistent with the GRU opponent embedding acting as a useful
regulariser for worst episodes, although the short horizon prevents a
replicated architectural conclusion.

\paragraph{CVaR reweighting.}
CVaR-MAPPO ($-5{,}131$, $\mathrm{CVaR} = -6{,}509$) vs MAPPO-MLP
($-3{,}937$, $\mathrm{CVaR} = -6{,}226$ at 200 episodes): at this training
horizon, episode-level return-tail reweighting by itself is less effective
than explicit operational-cost constraints.
This supports the paper's central design choice: tail-risk accounting is a
useful component of ACD\textsuperscript{3}, but SOC deployability is driven
primarily by the safety contract.

\paragraph{ACD\textsuperscript{3} component checks.}
The 30-episode AskHuman and deterministic G-CRP component runs
both reach
$P(\mathrm{viol}^{\mathrm{DT}})=0.100$, $\bar{c}^{\mathrm{DT}}=36.8$,
and $\bar{R}=-7{,}901$.
This is safer than the reported ACD\textsuperscript{3}-GAT result
on violation rate (0.100 vs 0.138), but the horizon is shorter and
the two variant metrics are identical at this summary level.
These short-horizon results show that the ACD\textsuperscript{3} switches can
operate within the safety-contract regime early in training.
They are component checks, not attribution evidence: because the AskHuman and
deterministic G-CRP rows collapse to identical aggregate metrics, they are not
used to claim separate causal effects for the two switches.
Their role is to verify that the override and deterministic G-CRP paths execute
within the same accounting framework as the replicated policies; causal
attribution would require per-step override reasons, proposed and executed
actions, $Q_{\mathrm{safe}}$ values, uncertainty estimates, and remaining-budget
state at the decision point.

\begin{figure*}[!t]
\centering
\includegraphics[width=0.95\textwidth]{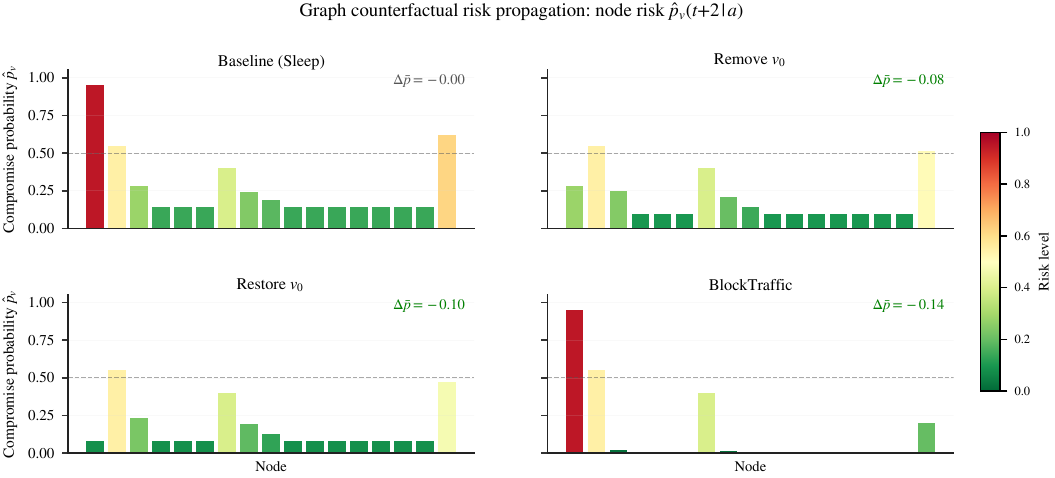}
\caption{Graph Counterfactual Risk Propagation (G-CRP) under alternative
  response actions.
  Each panel applies one counterfactual action to the same graph observation
  and propagates risk for two steps using the independent-cascade model in
  Equation~\eqref{eq:cascade}.  This figure illustrates the deterministic
  rollout-free shield model used by the reported policy runs; learned G-CRP is
  the supervised risk-estimation extension trained from logged trajectories.}
\label{fig:grp_physics}
\end{figure*}

\paragraph{Learned G-CRP fitting.}
Table~\ref{tab:grp_training} reports the supervised fitting result for
the learned G-CRP model.
The model was trained from six longer-horizon trajectory files,
using 6,000 sampled transitions from a 45,000-transition valid pool.
The loss decreased from 0.2638 at epoch 5 to 0.2606 at epoch 10, indicating
that the learned risk-propagation component can be fitted from the collected
safety-labelled trajectories.
This establishes the supervised risk-estimation path from the collected
trajectories.
The policy benchmark itself uses the deterministic G-CRP screen, keeping
policy performance claims tied to the action-screening rule used during the
reported evaluations.
Figure~\ref{fig:grp_physics} should therefore be read as a mechanism figure,
not as another benchmark row.
Each panel applies a different candidate response action to the same graph
belief state and propagates risk with the deterministic cascade in
Equation~\eqref{eq:cascade}.
The comparison illustrates how G-CRP changes the action-screening problem:
actions are not ranked only by immediate reward or immediate cost, but by the
post-action graph-risk field that would be handed back to the policy loop.
This is the counterfactual calculation behind $Q_{\mathrm{safe}}$ in
Equation~\eqref{eq:q_safe}.


\begin{table}[t]
\centering
\caption{Learned G-CRP fitting result.  The table reports the supervised
risk-estimation component trained from safety-labelled trajectories; policy
performance in the benchmark is tied to the deterministic G-CRP screen used
by the evaluated runs.}
\label{tab:grp_training}
\small
\begin{tabular}{l r}
\toprule
Quantity & Value \\
\midrule
Source trajectory files & 6 \\
Valid transition pool & 45,000 \\
Sampled transitions & 6,000 \\
Training epochs & 12 \\
Loss at epoch 5 & 0.2638 \\
Loss at epoch 10 & 0.2606 \\
\bottomrule
\end{tabular}
\end{table}


\paragraph{Temporal Contract Graph Shielding diagnostic.}
ACD\textsuperscript{3}-TCGS extends the action-screening question from
one-step graph risk to future contract risk.
The temporal model was trained on 40{,}000 length-8 sequences sampled from
180 safety-labelled episodes across 9 trajectory files.
On the held-out split, it achieves near-perfect discrimination for future
downtime-budget violation
($\mathrm{AUC}_{\mathrm{down}}=0.99998$,
Brier score $=0.00162$, calibration error $=0.00177$);
firewall-disruption and false-positive Restore risks show similarly high
discrimination
($\mathrm{AUC}_{\mathrm{fw}}=0.99827$ and
$\mathrm{AUC}_{\mathrm{fp}}=0.99995$).
At the diagnostic threshold $\epsilon=0.05$, the fitted model accepts
approximately 32.9\% of candidate histories, with an observed future
downtime-violation rate of 0.11\% among accepted samples.

\begin{figure*}[!t]
\centering
\includegraphics[width=0.95\textwidth]{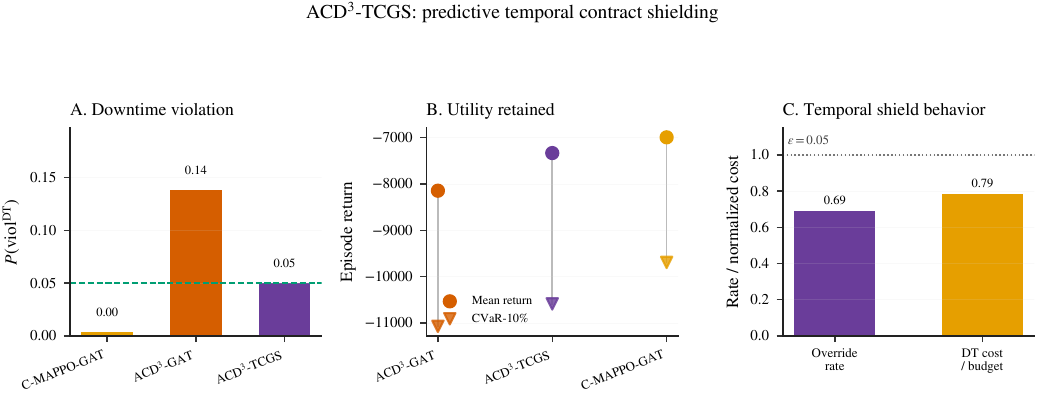}
\caption{Predictive temporal contract shielding diagnostic for
  ACD\textsuperscript{3}-TCGS.
  \textbf{(a)} The policy-level diagnostic compares downtime-budget violation
  for C-MAPPO-GAT, ACD\textsuperscript{3}-GAT, and the frozen
  ACD\textsuperscript{3}-GAT policy screened by the temporal contract-risk
  model.
  \textbf{(b)} Mean return and CVaR-10\% are reported on the same policies so
  the shield is interpreted as a safety--utility intervention rather than a
  violation-only post-processing step.
  \textbf{(c)} The final panel reports the deployability threshold
  $\epsilon=0.05$, override rate, and normalized downtime cost for the
  temporal shield.}
\label{fig:tcgs_policy_panel}
\end{figure*}

We also ran a frozen-policy ACD\textsuperscript{3}-GAT + TCGS screening
diagnostic. The completed frozen-policy diagnostic contains 100 episodes. It is included as single-seed policy-level diagnostic evidence. Mean downtime cost is 39.3, with downtime-violation rate 5.0\%; false-positive violation is 8.0\% and catastrophic rate is 0.0\%. The mean return is $-7{,}331$ with $\mathrm{CVaR}_{10\%}=-10{,}590$. The firewall-change budget remains the active weakness ($\bar c^{\mathrm{fw}}=105.2$, violation rate 31.0\%), so these results are used as temporal-shield evidence rather than as a headline claim that ACD\textsuperscript{3}-TCGS dominates the replicated policies.
Figure~\ref{fig:tcgs_policy_panel} visualizes this diagnostic frontier.



\subsection{Tail-Risk and Catastrophic Episodes}
\label{sec:tail_risk}

We define a \emph{catastrophic episode} as one in which the mean alert
level exceeds threshold $\theta_{\mathrm{alert}}{=}8$, corresponding to
coordinated multi-host compromise that threatens mission continuity.
Table~\ref{tab:benchmark} reports the catastrophic episode rate for all methods.

\textbf{CVaR reweighting alone is insufficient at this horizon.}
CVaR-MAPPO achieves a catastrophic rate of 0.090 -- \emph{higher} than
MAPPO-MLP (0.015 at 200 episodes) and MAPPO-GNN (0.010 at 100 episodes).
Its CVaR-10\% ($-6{,}509$) is also worse than the reported MAPPO-MLP and
MAPPO-GAT rows.
This confirms the finding from Section~\ref{sec:ablations}: at the reported
training horizon, the episodic diversity that CVaR needs to reweight has not
yet accumulated.

\textbf{Graph structure improves worst-tail return, not every tail proxy.}
MAPPO-GAT achieves the best CVaR-10\% among the reported unconstrained MAPPO
encoder rows ($-5{,}864$, catastrophic rate 0.035), outperforming MAPPO-MLP
($-6{,}226$, 0.015) and MAPPO-GNN ($-6{,}108$, 0.010) on worst-tail return.
The catastrophic-rate proxy does not improve in the same ordering:
MAPPO-GAT has a higher catastrophic rate than MAPPO-MLP and MAPPO-GNN.
Therefore, the graph attention improves the
worst-tail return metric in this benchmark, but it does not by itself remove catastrophic alert-threshold episodes.

\textbf{Constraints are the most reliable observed mechanism for reducing operational harm.}
C-MAPPO-GAT achieves catastrophic rate 0.000 and CVaR-10\% $= -9{,}694$.
The lower catastrophic rate reflects an important tradeoff: the Lagrangian
policy accepts lower mean return in exchange for bounded operational cost,
which removes alert-threshold catastrophic episodes in the C-MAPPO-GAT
evaluations.
The reported ACD\textsuperscript{3}-GAT result achieves catastrophic rate 0.003,
but its CVaR-10\% is $-11{,}074$, worse than C-MAPPO-GAT.
This boundary is informative, because composing graph encoding, CVaR reweighting, Lagrangian constraints, and shielding does not automatically dominate the cleaner constrained baseline. The 30-episode AskHuman and deterministic G-CRP component runs have lower catastrophic rate and lower downtime violation than the reported ACD\textsuperscript{3} result,
but their horizon is too short to identify a stable component effect.
Additional ACD\textsuperscript{3} replications and component ablations are
therefore needed to separate whether the penalty comes from override behavior, G-CRP screening, CVaR weighting, or insufficient training horizon.

\subsection{Robustness Stress Tests}
\label{sec:robustness_results}

The robustness studies explores a narrower question than the main benchmark. Once a policy has been trained, does the safety contract survive controlled changes in topology seed or adaptive Red-process behaviour? They are interpreted as stress tests rather than as a replacement for the replicated ID benchmark in Table~\ref{tab:benchmark}.

\paragraph{Topology-seed stress.}
On unseen topology seeds, the safety contrast remains intact.
Across 120 evaluation episodes, constrained MAPPO-GAT preserves
$P(\mathrm{viol}^{\mathrm{DT}})=0.000$, mean downtime cost $14.9$, and
catastrophic rate 0.000.
IPPO and MAPPO-GAT remain non-deployable under the same stress:
both have $P(\mathrm{viol}^{\mathrm{DT}})=1.000$, with mean downtime costs
223.6 and 345.7, respectively.
The topology shift does not harm MAPPO-GAT's reward
($\bar{R}=-3{,}517$ vs $-3{,}548$ on the matched ID evaluation). Thefore, the reward robustness alone is insufficient when the operational contract is exhausted in every episode. For constrained MAPPO-GAT, topology stress produces a stable safety profile ($\bar{R}=-7{,}286$, $\mathrm{CVaR}_{10\%}=-9{,}433$) while keeping the
MTTR budget satisfied. ACD\textsuperscript{3}-GAT is analysed on the matched-layout benchmark and adaptive Red-process stress test.
The topology-OOD comparison is restricted to policies whose action heads are already layout-stable under the evaluated unseen topology seeds.

\begin{figure*}[!t]
\centering
\includegraphics[width=0.95\textwidth]{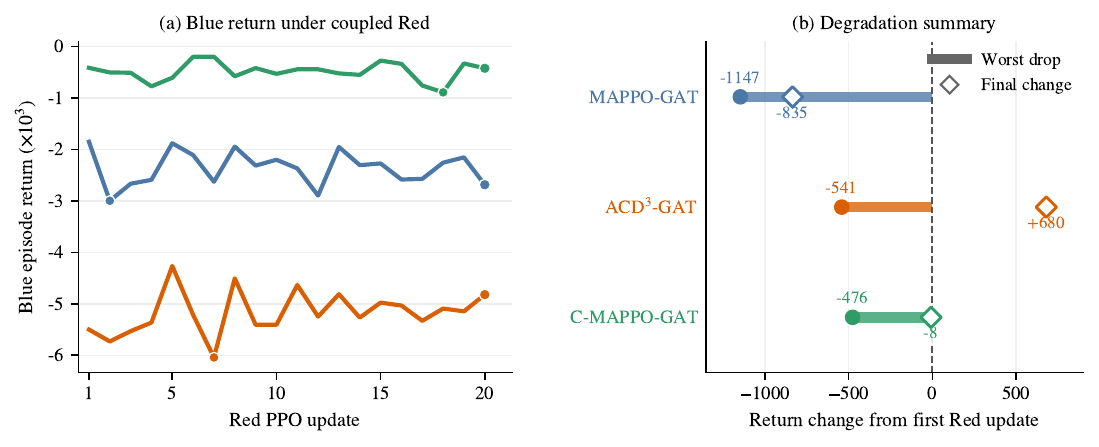}
\caption{Coupled adaptive Red-process stress test.
  Each curve evaluates a frozen Blue policy while a Red PPO learner adapts
  for 20 updates and sampled Red actions are injected into the CybORG
  transition loop.
  \textbf{(a)} Raw Blue-policy return under adaptation.
  \textbf{(b)} Worst and final change in Blue-policy return relative to the first Red-process
  update, which compares degradation despite different baseline return scales.
  The figure is a robustness stress test over one Blue-policy seed per
  method, not a multi-seed exploitability guarantee.}
\label{fig:exploitability}
\end{figure*}

\paragraph{Coupled adaptive Red process.}
Figure~\ref{fig:exploitability} reports the adaptive Red-process stress test. For each frozen Blue policy, a Red PPO learner is trained for 20 updates and its sampled actions are injected into the CybORG transition loop.
The raw-return panel is useful for scale, while the degradation panel compares how much each Blue policy worsens relative to its first Red-process update. MAPPO-GAT has the largest worst degradation, since its Blue return drops by 1,147 points at the worst update and finishes 835 points below its starting value.
The Constrained MAPPO-GAT is much more stable, with a worst degradation of
476 points and an end-of-run change of only $-8$ points.
ACD\textsuperscript{3}-GAT starts from a lower raw-return level, but its
worst degradation is 541 points and the curve recovers to finish 680 points
above its first update.
This supports a bounded robustness interpretation:
explicit safety machinery is associated with lower worst policy-return
degradation than reward-only MAPPO-GAT, while constrained MAPPO-GAT remains
the most reliable safety-contract policy among the evaluated methods.


\subsection{Synthesis}

Across the evaluated methods, reward-only learning is consistently
non-deployable under the stated SOC contract:
$P(\mathrm{viol}^{\mathrm{DT}})=1.000$ for every unconstrained learner, and the replicated IPPO/MAPPO rows use 314--355 downtime-cost units per episode against a budget of 50. Lagrangian safety machinery changes that operating regime. C-MAPPO-GAT reduces downtime cost to 15.5 and downtime-budget violation to 0.3\%, while ACD\textsuperscript{3}-GAT reduces mean downtime cost to 48.2 with a 13.8\% episode-violation rate.
The difference is central to the interpretation, because the C-MAPPO-GAT configuration introduced here is the most reliable safety-contract policy observed in the benchmark, whereas ACD\textsuperscript{3}-GAT is the broader architecture for integrating graph perception, safety contracts, tail-risk accounting, and counterfactual action screening.

The experiments also explain why simple alternatives are insufficient.
Naive reactive rules can perform worse than doing nothing
($\bar{R}\approx-8{,}124$ vs.\ $-6{,}792$ for Sleep) because they apply
restoration without the valid-action, observation-processing, and mission-policy discipline that made engineered CAGE~4 heuristics effective.
Reward-only MARL fails from the opposite direction, because it learns high-impact security interventions without learning when those interventions exhaust operational budgets.
Together, the results answer the central question of the paper:
\textit{autonomous network-security policies cannot be evaluated on reward alone, and operational safety contracts must be represented and optimized explicitly.} Future ACD\textsuperscript{3}-GAT evaluations should extend the component ablations and make the factorised target head layout-stable for topology generalisation.


\section{Discussion}
\label{sec:discussion}

\subsection{Evidence Across the Experiments}

The empirical evidence extends beyond a single algorithm comparison.
Together, the experiments characterise a difficult safety-contract MARL problem:
non-learning policies expose the cost of naive response rules; IPPO and MAPPO
variants show that reward-only learning discovers operationally harmful
interventions; graph encoders test whether network structure improves
representation and tail behaviour; constrained MAPPO-GAT isolates the effect
of explicit Lagrangian safety machinery; ACD\textsuperscript{3}-GAT integrates
graph perception, budget context, CVaR tail-risk accounting, override signals, and
counterfactual action screening; short AskHuman and deterministic G-CRP
evaluations examine component behaviour; topology-seed and coupled adaptive
Red-process stress tests
ask whether the safety contrast survives controlled robustness checks; and
the two 300-episode replications test whether the principal ordering persists
beyond the balanced 200-episode benchmark.
This breadth matters because the contrasts are scientifically informative:
graph attention improves the worst-tail return but not raw
return dominance; CVaR reweighting alone does not improve the tail at the
reported horizon; the integrated ACD\textsuperscript{3} policy reduces harm
while the constrained
safety machinery is the component that most consistently changes the
operational outcome.

\subsection{The Operational Safety Finding}

The most important result is not which method achieves the highest return,
but rather that \emph{every unconstrained method fails the safety contract
on every episode}.
Mean downtime costs of 311--430 against a budget of 50 mean that
unconstrained agents consume roughly 6--9$\times$ the allowable MTTR budget
per episode.
In a real SOC, this would manifest as hundreds of unnecessary host
reimages per shift, violation of SLA commitments, and systematic
availability degradation of the very assets being protected.

Constrained MAPPO-GAT reduces the mean downtime cost to 15.5 ($<B_{\mathrm{down}}$)
and the violation rate to 0.3\% across three 200-episode seeds.
ACD\textsuperscript{3}-GAT also brings mean
downtime cost below budget (48.2), with a 13.8\% episode-violation rate,
so the integrated architecture occupies a different point in the design space:
it demonstrates the full safety-contract stack, whereas C-MAPPO-GAT is the
new controlled constrained configuration that most reliably satisfies the
SOC contract in the present benchmark.
The two additional 300-episode replications strengthen rather than soften
this interpretation, since constrained MAPPO-GAT remains near-zero on downtime violations (0.7\%), reward-only MAPPO-GAT remains at 100\% violation, and ACD\textsuperscript{3}-GAT remains close to budget on average while still violating in 14.3\% of episodes.
C-MAPPO-GAT incurs a return penalty of approximately 3,050 points relative
to the highest-return unconstrained learner, MAPPO-MLP
($-6{,}992$ vs.\ $-3{,}937$).
ACD\textsuperscript{3}-GAT incurs approximately 4,200 points
($-8{,}144$ vs.\ $-3{,}937$), consistent with the broader set of safety
signals active in the integrated safety-contract stack.
We argue that this tradeoff is not only acceptable but \emph{necessary}:
a policy that achieves high security reward while violating operational
budgets is not deployable, regardless of its average performance.

\subsection{Role of ACD\textsuperscript{3}-GAT}

ACD\textsuperscript{3}-GAT functions as the integrated method architecture, while C-MAPPO-GAT is the strongest compliance row in the current benchmark.
Its role is methodological, since it specifies a general safety-contract architecture with graph-structured perception over network entities, Lagrangian cost learning, explicit budget context, counterfactual action screening, tail-risk accounting, and override signals.
C-MAPPO-GAT is the strongest safety-contract configuration because it
isolates the new combination of MAPPO, a GAT encoder, and Lagrangian
operational-cost control, satisfying the downtime contract most reliably in the reported evaluations. ACD\textsuperscript{3}-GAT is the extensible architecture: it unifies the reusable components needed when a response policy must reason over changing topology,
opposing-process adaptation, uncertain action consequences, and human-governed operational budgets. The 13.8\% downtime-violation rate therefore identifies the key stabilization target for the integrated policy while preserving the broader method
contribution.

\subsection{Failure Mechanism of Unconstrained MARL}

The mechanism is clear from the cost breakdown.
Unconstrained MAPPO-MLP issues 316 downtime-cost units per episode (vs.\ budget
50); IPPO issues 314 and MAPPO-GAT issues 355.
The rule-based heuristic, despite being the lowest-returning baseline,
incurs 117 downtime-cost units because it applies Restore selectively.
Random exploration averages 424 downtime-cost units---worse than the
constrained agent because it occasionally selects Restore by chance.

This refines the intuition from the published CAGE~4 analyses:
the competition heuristics succeeded because they encoded observation handling,
invalid-action avoidance, mission-phase firewall discipline, and selective
response policies.
Our naive rule baseline lacks that structure and becomes operationally harmful.
Unconstrained RL fails from the other side, because it finds Restore and BlockTraffic attractive because they genuinely reduce the CAGE Red process's presence in the short term, but the Lagrangian penalties in our framework counterbalance that exploitable pattern with explicit operational budgets.

\subsection{Graph Encoder Evidence}

In the benchmark, MAPPO-MLP at 200 episodes ($-3{,}937$)
slightly outperforms MAPPO-GAT across 600 episodes ($-3{,}979$),
while MAPPO-GNN remains
available at 100 episodes ($-4{,}387$).
This is consistent with slower convergence of graph encoders due to their
larger parameter count and the need to learn structural attention weights.
These data support a specific graph-encoder conclusion rather than a broad
raw-return dominance conclusion.
Critically, MAPPO-GAT achieves the best CVaR-10\% among the reported
unconstrained MAPPO encoder rows ($-5{,}864$), and MAPPO-GNN also improves
the tail relative to MAPPO-MLP,
suggesting that graph structure is presently a tail-risk and interpretability
result rather than a mean-return result.

\subsection{Scope of Evidence and Transfer}

\paragraph{Claim boundary.}
This study is intentionally scoped to safety-contract evaluation of autonomous
network-security response in CAGE-4.
The primary inferential target is the contrast between reward-only MARL and
explicit operational-cost control, not return-based ranking,
broad topology generalisation, or factorial attribution of every
ACD\textsuperscript{3} component.
Three replicated seeds support stability of the safety contrast for the core
methods, and rare violation rates are reported with raw episode counts rather
than as formal chance-constraint certificates.
Robustness results are controlled stress tests under the same cost accounting.
Larger seed counts, hierarchical task decomposition, layout-stable topology
evaluation for ACD\textsuperscript{3}-GAT, and head-to-head comparison with
recent graph-RL and hierarchical MARL systems under a shared safety logger are
complementary follow-on work rather than prerequisites for the central
deployability finding.

\paragraph{Simulator-grounded evidence.}
The study is intentionally grounded in CybORG/CAGE-4, where actions,
observations, CAGE Red-process behaviour, and mission scoring are controlled and
reproducible.
The cost proxies ($c_t^{\mathrm{down}}, c_t^{\mathrm{fw}}, c_t^{\mathrm{fp}}$)
map the simulator primitives onto SOC-relevant governance quantities:
service recovery burden, firewall-change burden, and false-positive response
burden.
This makes the safety contract auditable inside the benchmark and gives
practitioners a clear template for replacing the proxies with organisation-
specific measurements in a live range, SOAR platform, or enterprise digital
twin.

\paragraph{Robustness and attribution scope.}
The replicated core comparison covers IPPO, MAPPO-GAT, and
constrained MAPPO-GAT, and ACD\textsuperscript{3}-GAT at three
200-episode seeds.
Two additional 300-episode replications cover MAPPO-GAT,
constrained MAPPO-GAT, and ACD\textsuperscript{3}-GAT, and are reported as a
longer-horizon replication check rather than as a replacement for the balanced
core table.
The topology-seed stress test covers IPPO, MAPPO-GAT, and
constrained MAPPO-GAT; the coupled adaptive Red-process stress test covers MAPPO-GAT,
constrained MAPPO-GAT, and ACD\textsuperscript{3}-GAT.
These choices define the evidence that a replicated
safety-contract comparison, a longer-horizon check of the principal ordering, and controlled robustness probes. The 30-episode AskHuman and deterministic G-CRP component rows are therefore
treated only as execution-path checks; their identical aggregate summaries do not identify which switch caused the observed behaviour.
The next attribution layer is naturally factorial: C-MAPPO-MLP, budget-context only, hard-shield only, Lagrangian-only, CVaR-only, opponent-context, and G-CRP-screening variants can be used to decompose the integrated ACD\textsuperscript{3} policy into its causal components.
That decomposition is an extension of the present benchmark rather than a
prerequisite for the core result that explicit safety machinery changes the operational regime.

\paragraph{Temporal and operational transfer.}
The 500-step, $\gamma{=}0.99$ formulation provides a controlled horizon for
comparing policies under identical mission dynamics.
Longer incident campaigns, analyst workflows, ticketing delays, and
organisation-specific change-control rules can be incorporated by changing the
cost functions and episode horizon while preserving the same constrained
Dec-POMDP and safety-contract machinery.
ACD\textsuperscript{3}-TCGS is the first step in that direction, where rather than screening only the immediate proxy cost or a
two-hop deterministic graph cascade, it learns from trajectory histories to predict future budget exhaustion before the action is executed.
The current evidence is frozen-policy temporal-shield evidence, not an end-to-end retrained TCGS policy result.
That boundary is important, but the result is still informative. It shows that the safety-labelled trajectories produced by the benchmark contain enough temporal structure to support predictive contract-risk screening.

\paragraph{Proxy cost signals.}
The false-positive Restore proxy (Restore with no active alerts) is
defined from visible alert evidence because the Blue response agents act under partial
observability.
This choice matches the operational question faced by SOC automation:
whether an action was justified by available evidence at decision time.
In a deployment audit, the same definition can be paired with ground-truth
incident labels when post-incident forensics are available.

\paragraph{Budget and chance-constraint sensitivity.}
The Lagrangian update controls expected episode cost, not violation
probability directly.
The empirical violation rate is therefore a measured outcome rather than a
formal chance-constraint guarantee.  Changing
$B_{\mathrm{down}}$, $B_{\mathrm{fw}}$, or $B_{\mathrm{fp}}$ would change both
the feasible policy set and the shield activation pattern.
The CAGE-4 budgets used here are therefore best read as reproducible
stress-test thresholds; operational adoption would instantiate the same
method with locally chosen service, change-management, and analyst-capacity
budgets.

\paragraph{Interpretation boundary.}
The safety contract is an evaluation and training discipline inside CAGE-4:
it exposes when a policy exhausts MTTR, firewall-change, or false-positive
budgets and provides the decision record needed for audit.
Graph attention is reported as topology-aware representation learning, and
G-CRP is reported as the deterministic counterfactual risk screen used by the
shield.
The learned G-CRP model is included in the replication package as the next
risk-estimation component in the same framework; the reported policy
performance is tied to the deterministic screen unless explicitly stated
otherwise.

\paragraph{Prior-work references.}
The published CC4 heuristic and LLM scores are important context, but they are
not direct safety-contract baselines.
They use different reported outputs and do not expose the per-action
Restore, Block/Allow, and false-positive traces required by our cost metrics.
We therefore use them to calibrate the reward scale, while the main claims are
restricted to methods evaluated with the same safety-labelled protocol.

\subsection{Implications for Autonomous SOC Deployment}

Read as an expert-system result, the framework provides a deployable
decision-support pattern rather than a claim of autonomous production
readiness, because we encode SOC governance as budgets, train and evaluate policies against those budgets, log every executed action and cost proxy, and route budget exhaustion or low-confidence states to human governance.
Our results suggest three practical principles for deploying such agents:
\begin{enumerate}
\item \textbf{Define operational budgets before training.}
  Unconstrained training consistently produces operationally harmful policies.
  Budget values ($B_k$) should be set by SOC governance and treated as
  explicit violation constraints with audit metrics, not as hidden soft
  reward preferences.
  The CAGE-4 values used here are stress-test thresholds; a SOC deployment
  should tune them from service-level objectives, change-management policy,
  and analyst capacity, then rerun the same safety-labelled evaluation.

\item \textbf{Validate robustness after satisfying the safety contract.}
  Fixed scripted Red processes can overestimate policy robustness, so adaptive
  opposing-policy and topology-shift studies should follow the ID safety benchmark
  before any deployment claim.
  In the reported stress tests, constrained MAPPO-GAT preserves zero downtime
  violations under topology-seed shift and degrades less than reward-only
  MAPPO-GAT under the coupled adaptive Red-process stress test.

\item \textbf{Support human override.}
  The ACD$^3$-GAT override mechanism (triggered by low confidence,
  budget exhaustion, or OOD states) provides a governance layer that
  maintains human authority over high-stakes decisions.
  AskHuman variants measure the cost of that governance layer and connect the
  learned policy to a SOC approval workflow.
\end{enumerate}


\section{Conclusion}
\label{sec:conclusion}

Autonomous network security agents trained without explicit operational
constraints systematically violate SOC budgets in the evaluated CAGE~4 setting,
creating downtime costs roughly 6--9$\times$ the allowable MTTR budget.
This is not merely a reward-engineering failure; it reflects a structural
mismatch between scalar security reward and operational deployability.

ACD$^3$-GAT addresses this through a unified safety-contract design:
Lagrangian-constrained multi-agent PPO that optimizes under MTTR,
false-positive, and firewall-policy budgets; graph encoders for structured
host--subnet observations; a factorised type--target action policy; a
multi-objective training objective for reward, operational cost, and
tail-risk accounting; and a budget-aware counterfactual shield.
On CAGE Challenge~4, the C-MAPPO-GAT constrained configuration introduced in
this work achieves a 95--96\% reduction in
operational harm while remaining within budget 99.7\% of episodes---a
qualitative shift from policy that is operationally harmful to one that
is operationally auditable.
Two additional 300-episode replications preserve the same ordering:
constrained MAPPO-GAT remains near-zero on downtime violations, while
reward-only MAPPO-GAT continues to violate the downtime budget in every
episode.
ACD$^3$-GAT reduces mean downtime harm but still
violates the budget in 13.8\% of episodes.
This identifies a clear division between contribution types:
C-MAPPO-GAT is the strongest compliance configuration in the reported
benchmark and demonstrates that the MAPPO/GAT/Lagrangian composition is a
powerful safety-contract baseline, while ACD$^3$-GAT is the general
architecture for combining graph
perception, safety contracts, tail-risk accounting, opponent context, and
counterfactual action screening.
Topology-seed and coupled adaptive Red-process stress tests reinforce
the same conclusion, which is that explicit safety machinery, not reward alone, determines whether an autonomous response policy remains operationally auditable under stress.

The core finding has direct implications for SOC governance: autonomous
network-security agents must carry explicit operational constraints before they can be considered for deployment. More broadly, ACD$^3$-GAT treats network security response as one instance of a wider class of safety-contract MARL problems. For instance, multiple agents act over structured entities, local actions have operational side effects, and deployment requires respecting budgets that are not reducible to reward. The anonymised replication package provides safety-labelled trajectories, trained checkpoints, evaluation metadata, and robustness protocols so that the reported claims can be independently regenerated and extended. The ACD\textsuperscript{3}-TCGS diagnostic shows how those trajectories can also train a temporal contract-risk shield that predicts future budget exhaustion before executing a proposed action.
The natural next research agenda is to instantiate the same safety-contract
framework at larger scale, use layout-stable target heads for broader topology
transfer, evaluate learned Red-process variants in the main robustness loop, integrate
the learned G-CRP risk model into policy evaluation, and pair the safety
contract with temporal graph memory or asynchronous cyber-range simulators
when moving from CAGE-style episodes to richer SOC telemetry
~\citep{rossi2020temporal}.  Continuous-time cyber-range evaluation provides
the complementary simulator direction~\citep{jankowski2026netforge}.

\FloatBarrier


\section*{CRediT Author Contribution Statement}

\textbf{Jose Luis Silva:} The author was responsible for the conceptualization and methodology of the study, developed the software, conducted the formal analysis and investigation, curated the data, prepared the original draft, reviewed and edited the manuscript, and produced the visualizations.

\section*{Declaration of Competing Interest}
The authors declare no competing financial or personal interests.

\section*{Data Availability}
Software, safety-labelled trajectories, trained model checkpoints, and robustness-evaluation metadata are available from the corresponding author upon request.

\bibliographystyle{cas-model2-names}
\bibliography{references}

\end{document}